%% file: mainarxiv.tex
\documentclass[12pt,a4]{article}

\usepackage{amsthm,amsmath,latexsym,amssymb,amsfonts,amscd}
\usepackage{graphics,lscape,fancyhdr,array,stmaryrd,euscript}
\usepackage{empheq}
\usepackage{dsfont}
\usepackage{verbatim}
\usepackage{relsize,slashed}
\numberwithin{equation}{section}
\usepackage{hyperref}
\input{YoungDiagrams.tex}

\usepackage{setspace}

 \usepackage[numbers,sort&compress]{natbib}
 \usepackage[nottoc]{tocbibind}
\newcommand{\pl}{\partial}

\newcommand{\ga}{\alpha}
\newcommand{\gb}{\beta}
\newcommand{\gc}{\gamma}

\newcommand{\gad}{{\dot{\alpha}}}
\newcommand{\gbd}{{\dot{\beta}}}
\newcommand{\gdd}{{\dot{\gamma}}}

\newcommand{\mm}{{\ensuremath{\underline{m}}}}

\newcommand{\fud}[2]{{}^{#1}{}_{#2}\,}

\newcommand{\Tr}{{\mathrm{Tr}\,}}

\newcommand{\hs}{\mathfrak{hs}}

\renewcommand{\d}{\partial}
\newcommand{\dd}{\partial}    
\newcommand{\bref}[1]{\textbf{\ref{#1}}}
\newcommand{\Dh}{\mathrm{D_{h}}}
\newcommand{\gh}[1]{\mathrm{gh}(#1)}

\newcommand{\ffrac}[2]{\raisebox{.5pt}%
  {\footnotesize$\displaystyle\frac{#1}{#2}$}\kern1pt}
\newcommand{\dl}[1]{\mathchoice{\ffrac{\dd}{\dd #1}}{\frac{\dd}{\dd #1}}{\ffrac{\dd}{\dd #1}}{\ffrac{\dd}{\dd #1}}}
\newcommand{\algA}{\mathcal{A}}
\newcommand{\commut}[2]{[#1{,}\,#2]}
\newcommand{\algg}{\Liealg{g}}
\newcommand{\Liealg}{\mathfrak} 
\def\cF{\mathcal{F}}
\newcommand{\inner}[2]{\langle #1{,}\,#2\rangle}
\newcommand{\p}[1]{|#1|}

\begin{document}
\pagenumbering{gobble}
\hfill
\vspace{-1.5cm}
\vskip 0.05\textheight
\begin{center}
{\Large\bfseries 

On matter-free Higher Spin Gravities in $\boldsymbol{3d}$:\\
\rule{0pt}{20pt}(partially)-massless fields and general structure}

\vspace{0.4cm}

\vskip 0.03\textheight

Maxim \textsc{Grigoriev}${}^{a,b}$, Karapet \textsc{Mkrtchyan}${}^c$ and Evgeny \textsc{Skvortsov}${}^{a,d}$

\vskip 0.03\textheight

\vspace{5pt}
{\em
$^a$ Lebedev Institute of Physics, \\
Leninsky ave. 53, 119991 Moscow, Russia\\

\vspace{5pt}

$^b$
Institute for Theoretical and Mathematical Physics,\\
Lomonosov Moscow State University, 119991 Moscow, Russia
\\
\vspace{5pt}
$^c$ Scuola Normale Superiore and INFN, \\ Piazza dei Cavalieri 7, 56126 Pisa, Italy\\
\vspace{5pt}
$^d$  Albert Einstein Institute, \\
Am M\"{u}hlenberg 1, D-14476, Potsdam-Golm, Germany
}
\end{center}

\vskip 0.02\textheight

\begin{abstract}
We study the problem of interacting theories with (partially)-massless and conformal higher spin fields without matter in three dimensions. A new class of theories that have partially-massless fields is found, which significantly extends the well-known class of purely massless theories. More generally, it is proved that the complete theory has to have a form of the flatness condition for a connection of a Lie algebra, which, provided there is a non-degenerate invariant bilinear form, can be derived from the Chern-Simons action. We also point out the existence of higher spin theories without the dynamical graviton in the spectrum. As an application of a more general statement that the frame-like formulation can be systematically constructed starting from the metric one by employing a combination of the local BRST cohomology technique and the parent formulation approach, we also obtain an explicit uplift of any given metric-like vertex to its frame-like counterpart. This procedure is valid for general gauge theories while in the case of higher spin fields in d-dimensional Minkowski space one can even use as a starting point metric-like vertices in the transverse-traceless gauge. In particular, this gives the fully off-shell lift for transverse-traceless vertices.

\end{abstract}
\newpage
\tableofcontents
\newpage
\section{Introduction and main results}
\pagenumbering{arabic}
\setcounter{page}{2}
Massless higher spin fields, as well as the graviton, do not have any propagating degrees of freedom in three dimensions. Therefore, the problem of constructing interactions of higher spin fields is subtle to formulate in $3d$. For example, within the light-cone approach \cite{Bengtsson:1983pg,Bengtsson:1986kh,Metsaev:2005ar}, which operates with local physical degrees of freedom, the problem is clearly empty. Nevertheless, as in the case of gravity, it makes sense to pick an off-shell gauge-invariant formulation of free fields inherited from higher dimensions and look for its nonlinear completion. Once the light-cone approach is out, there are two other common off-shell formulations: metric-like and frame-like. 

The metric-like formulation operates with a higher spin generalization of the metric tensor, $\Phi^{a_1...a_s}(x)$. The frame-like formulation leads in $3d$ to one-form connections $A^{a_1...a_{s_1}}_\mm \, dx^\mm$. The problem of interactions boils down to constructing gauge invariant actions in terms of one or the other set of variables. Even though these two approaches are directed to solve the same problem -- constructing an action for higher-spin gravity, they have developed independently from each other and their relation remains little explored beyond free theory, see however \cite{Grigoriev:2010ic,Grigoriev:2012xg,Campoleoni:2012hp,Fredenhagen:2014oua}.

In the metric-like formulation the problem of interactions, so-called Fronsdal programme, has advanced significantly during the last decades. There is a lot of results on the general structure of perturbative interaction vertices available in the literature  \cite{Berends:1984rq,Boulanger:2006gr,Francia:2007qt,Fotopoulos:2008ka, Zinoviev:2008ck, Boulanger:2008tg, Manvelyan:2010wp} ranging from the complete classification of cubic vertices in flat space \cite{Manvelyan:2010jr} (see also \cite{Sagnotti:2010at,Fotopoulos:2010ay,Manvelyan:2010je,Metsaev:2012uy,Fredenhagen:2019lsz}) to its extension to $(A)dS$ space \cite{Joung:2011ww,Francia:2016weg} (see also \cite{Bekaert:2014cea,Sleight:2016dba,Joung:2019wbl}) that incorporates partially-massless fields. Cubic interactions of conformal fields were also studied \cite{Metsaev:2016rpa}. Quite independently of the here-above results, there is also a handful of papers devoted to the frame-like approach \cite{Fradkin:1986qy,Lopatin:1987hz,Alkalaev:2002rq,Alkalaev:2010af,Boulanger:2011qt,Zinoviev:2010cr,Boulanger:2012dx}. 

More specifically, in three dimension, the classification of interaction vertices of massless higher spin fields has been worked out recently in \cite{Mkrtchyan:2017ixk,Kessel:2018ugi,Fredenhagen:2018guf,Fredenhagen:2019hvb}. At the same time there is a large number of concrete examples of theories with massless higher spin fields \cite{Blencowe:1988gj,Bergshoeff:1989ns,Campoleoni:2010zq,Henneaux:2010xg} and of theories with conformal higher spin fields  \cite{Pope:1989vj,Fradkin:1989xt,Grigoriev:2019xmp}, both classes having been constructed within the frame-like approach as Chern-Simons theories. The question of whether all theories are of Chern-Simons type and whether there are theories with partially-massless fields has remained open.

It is known that frame-like actions can be rewritten in terms of the metric-like fields, but the opposite is more complicated.  
In principle, the frame-like form of a given metric-like vertex can be obtained by employing the  Lagrangian parent formulation approach~\cite{Grigoriev:2010ic,Grigoriev:2012xg} which, among other applications, allows to systematically reformulate a Lagrangian gauge theory in the frame-like form. However, this does not directly give a concise and handful procedure to obtain frame-like vertices. One of the goals of the present work is to propose such a procedure and explicitly demonstrate how it works in the case of higher spin theories in 3d. 

The general framework to address problems of this sort in the context of local gauge field theories is known by now and is based on the combination of Batalin-Vilkovisky formalism~\cite{Batalin:1981jr,Batalin:1983wj} with the geometric theory of PDE (partial differential equations)~\cite{Vinogradov1981,Anderson1991,Dickey:1991xa,Olver:1993,vinogradov:2001,Krasil'shchik:2010ij}. This gives a powerful approach of local BRST cohomology~\cite{DuboisViolette:1985jb,Barnich:1995db,Piguet:1995er,Barnich:2000zw} and allows to reformulate the problem of gauge theories deformations and analysis of vertices as a standard deformation theory where the relevant cohomology is a local BRST cohomology~\cite{Barnich:1993vg}. Furthermore, in this approach one can introduce a general notion of equivalence of local gauge field theories~\cite{Dresse:1990dj} that covers theories related by elimination of auxiliary fields as well as by elimination of Stueckelberg fields. This notion extends~\cite{Barnich:2004cr}, see also~\cite{Barnich:2010sw,Grigoriev:2012xg}, to systems defined at the level of equations of motion in which case it also extends to a more general geometrical setting~\cite{Grigoriev:2019ojp}. 

For instance, starting from the BRST complex of the metric-like theory one can construct its  equivalent form, often called minimal model,\footnote{Note that in the literature on local BRST cohomology the term ``minimal model'' was not used extensively. The relevance of these formulations was realized by F.~Brandt who called it formulation in terms of generalized curvatures and connections.} obtained by elimination of the maximal amount of contractible pairs of the total differential. Such minimal formulations are known to be very useful in studying local BRST cohomology~\cite{Barnich:1995ap,Brandt:1996mh,Brandt:1997iu,Brandt:2001tg}. Moreover,  it was shown that the minimal model of a BRST complex actually encodes the frame-like formulation of the theory through the so-called parent formulation construction~\cite{Barnich:2004cr,Barnich:2010sw} (see also \cite{Grigoriev:2010ic,Grigoriev:2012xg} for a Lagrangian version). At the level of equations of motion the latter can be explicitly read-off from the minimal model as a generalized AKSZ-type sigma model. As we demonstrate in this work, this allows one to explicitly construct frame-like vertices starting from a representative of the respective cohomology class in the minimal model of the BRST complex and hence gives a systematic way to construct the frame-like counterpart of a given metric-like vertex.\footnote{Note that this can also be inferred from the Lagrangian parent formulation. For instance applying the procedure~\cite{Grigoriev:2010ic,Grigoriev:2012xg} to the metric like Lagrangian (for e.g. Fronsdal fields) perturbed by a cubic vertex one in principle arrives at the frame-like Lagrangian perturbed by the frame-like version of the vertex. However, this is not very efficient in this context as it requires extra variables.}  More precisely, the procedure amounts to first constructing a completion of the vertex to a cocycle (understood as a $d$-form on the jet-space) of the total BRST differential $\tilde s=\Dh+s$, involving the total de Rham differential $\Dh$. Then one reduces the cocycle to the minimal model, which can be understood as a surface in the original jet-space. Finally, one constructs the vertex by evaluating the cocycle on a field configuration.

The advantage of the frame-like formulation becomes overwhelming in three dimensions. While one can write down a lot of expressions that are nonlinear in $\Phi^{a_1...a_s}(x)$ and have derivatives contracted in various ways, there is a unique nonlinear functional of type $A\wedge A\wedge A$ that is a three-form. Remarkably, in the frame-like language the weak field expansion stops at cubic terms, but it is an infinite series in terms of metric-like fields. Given the equivalence between the two approaches, we can stick to the frame-like one as the simpler one to solve the problem of constructing higher spin theories without matter in three dimensions.

We aim to construct and describe all higher spin gravities in three dimensions whose off-shell field content consists of massless, partially-massless or conformal fields that have no on-shell propagating degrees of freedom.\footnote{\label{nomatter}Note that for $s=1$ and also for the maximal depth partially-massless fields one can choose between two on-shell descriptions, one with a propagating degree of freedom and another with none. We always consider the second option. Matter fields are also excluded. } Not surprisingly, all these theories turn out to have the Chern-Simons/flat connection form for an appropriate choice of the gauge algebra. In this context, the algebras are called higher spin algebras even though all of them emerge from the endomorphism algebras, $\mathrm{End}(V)$, for an appropriate $V$. We construct a large class of such algebras, which, in particular, leads to new theories.

Higher spin gravities in three dimension have been extensively studied in view of their holographic applications \cite{Blencowe:1988gj,Bergshoeff:1989ns,Campoleoni:2010zq,Henneaux:2010xg}. The Einstein-Hilbert action can be rewritten as the Chern-Simons action for $sl_2\oplus sl_2$ \cite{Achucarro:1987vz,Witten:1988hc}. The starting point for the higher spin generalization was to replace $sl_2$ with any bigger algebra $\mathfrak{g}\supset sl_2$ and write down the Chern-Simons action for $\mathfrak{g}\oplus \mathfrak{g}$. An implicit, but important ingredient here is an embedding of $sl_2$ into $\mathfrak{g}$ and the fact that $\mathfrak{g}$ is an $sl_2$-module. The decomposition of $\mathfrak{g}$ into $sl_2$-modules lists out the spectrum of massless fields, the rule being that a spin-$s$ field corresponds to two connections that take values in the dimension $(2s-1)$ irreducible $sl_2$-module $V_{s-1}$:
\begin{align}
    \Omega^{\ga(2s-2)}\,,\qquad\qquad \Omega^{\gad(2s-2)}
\end{align}
Here we used the language of spin-tensors, $\ga,\gad,...=1,2$, and we also made it clear that one connection is a module of the first $sl_2$ and the other is a module of the second $sl_2$. This construction yields a large class of theories with massless higher spin fields and the graviton. However, it has not been clear if all possible theories are covered by such a construction. The systematic study of higher spin interactions in $3d$ within the metric-like formalism and Noether procedure has been carried out only recently in \cite{Mkrtchyan:2017ixk,Kessel:2018ugi,Fredenhagen:2018guf,Fredenhagen:2019hvb}.

Another interesting problem is to construct higher spin theories with partially-massless fields. Indeed, partially-massless higher spin fields, like the massless ones, have no local degrees of freedom in three dimensions, c.f. footnote \ref{nomatter}. The frame-like formulation has been worked out in \cite{Skvortsov:2006at} and studied in three dimensions in \cite{Buchbinder:2012bz}. While \cite{Skvortsov:2006at} operates with many frame-like fields of the Lorentz algebra (but a single connection of the anti-de Sitter algebra), it turns out that they correspond to just two representations of $sl_2\oplus sl_2$ \cite{Gwak:2015jdo}: 
\begin{align}
    \Omega^{\ga(2s-t-1),\gad(t-1)}\,,\qquad \Omega^{\ga(t-1),\gad(2s-t-1)}\,,
\end{align}
where $t$ is the depth of partially-masslessness and $t=1$ corresponds to massless fields. The massless case is somewhat degenerate and does not allow one to see that the connection is charged, in general, with respect to both $sl_2$ subalgebras. This observation solves the puzzle and allows us to construct a new class of theories with partially-massless fields. It would be very interesting to study their holographic applications along the lines of \cite{Campoleoni:2010zq,Henneaux:2010xg,Afshar:2011qw}.

The case of $3d$ conformal higher spin fields can be treated analogously, the only difference being is that we should be looking for a higher spin extension of the conformal algebra $so(3,2)$ or $so(4,1)$ rather than of $sl_2\oplus sl_2$. 

To summarize, our results are as follows.

\vspace{-0.4cm}
\begin{itemize}\setlength\itemsep{-3pt}
    \item we construct a new class of theories with partially-massless fields;
   
    \item we give an explicit construction for the frame-like vertex in terms of a metric-like one. It is also shown that the transverse-traceless gauge, which is usually used in the metric-like language, can always be lifted. The latter two statements are true for any spacetime dimension;

    \item we prove that all diffeomorphism invariant higher spin gravities without propagating matter and involving massless, partially-massless and conformal higher spin fields have to have the form of the flatness condition for a certain higher spin algebra;

    \item provided the algebra admits a non-degenerate invariant bilinear form, the equations can be obtained from the Chern-Simons action. This completes the Noether procedure in $3d$ for the matter-free higher spin theories;\footnote{The result extends the already known $3d$ massless  \cite{Blencowe:1988gj,Bergshoeff:1989ns,Campoleoni:2010zq,Henneaux:2010xg} and conformal theories \cite{Pope:1989vj,Fradkin:1989xt,Grigoriev:2019xmp}. Other complete solutions of the Noether procedure include $4d$ conformal  \cite{Segal:2002gd,Tseytlin:2002gz,Bekaert:2010ky} and $4d$ Chiral \cite{Metsaev:1991mt,Metsaev:1991nb,Ponomarev:2016lrm} higher spin theories. }
    
    \item as a by-product, we classify vertices for (partially)-massless fields;
    
    \item we point out an existence of higher spin theories whose spectrum does not contain the graviton, i.e. they are formulated on a fixed gravitational background, and are not diffeomorphism invariant. This phenomenon has also some analogs in higher dimensions \cite{Gunaydin:2016amv,Giombi:2016pvg} for the Type-B,C theories.
    
\end{itemize}
We begin in section \bref{sec:freefields} with a short description of the metric-like and frame-like formulations for free fields. A large class of theories, including the new class with partially-massless fields, can be found in section \bref{sec:hsgravity}. The detailed discussion of the relation between frame-like and metric-like languages is in section \bref{sec:equivalence}, where we also complete the Noether procedure and prove the theories in question to have the Chern-Simons form as was anticipated in \cite{Fredenhagen:2019hvb} for massless higher spin fields.

\section{Free fields: metric-like vs. frame-like}
\label{sec:freefields}
We briefly describe three classes of (higher spin) fields for which we would like to construct interacting theories. There are two standard choices of field variables: metric-like fields and frame-like fields. While the former is the most canonical choice, it is the latter that can be efficiently pushed to the interacting level for three-dimensional theories without propagating degrees of freedom in the bulk.

\paragraph{Massless fields.} It is customary to begin with the Fronsdal approach \cite{Fronsdal:1978rb}, where a spin-$s$ field is a symmetric rank-$s$ tensor $\Phi^{a(s)}\equiv\Phi^{a_1...a_s}(x)$ that is subject to the following gauge symmetry\footnote{Indices $a,b,c,...=0,...,d-1$ are indices of the local Lorentz algebra, $so(d-1,1)$. They can be converted to world indices with the help of the dreibein $h^a_\mu$. Everything takes place in $3d$ anti-de Sitter space with dreibein $h^a_\mu$ and spin-connection $\varpi^{a,b}_\mu$. The fiber indices are raised, lowered and contracted with the flat metric $\eta_{ab}$ and we never have to use the anti-de Sitter metric explicitly.  Our shorthand notation implies that $a(s)$ denotes a group of symmetric indices $a_1...a_s$. Nevertheless, most of the discussion below is valid in the Minkowski space as well. The indices to be symmetrized are all denoted by the same letter, i.e. $\nabla^a \xi^{a(s-1)}$ unfolds to $s$ terms. }  
\begin{align}
   \delta\Phi^{a(s)}=\nabla^a \xi^{a(s-1)}\equiv \nabla^{a_1} \xi^{a_2...a_s}+\text{permutations}\,.
\end{align}
The field is double-traceless, $\Phi\fud{a(s-4)bc}{bc}=0$, and the gauge parameter $\xi^{a(s-1)}$ is traceless, $\xi\fud{a(s-3)b}{b}=0$. The gauge-invariant equations of motion read
\begin{align}\label{adsfron}
\square \Phi^{a(s)}-\nabla^a \nabla_m\Phi^{m a(s-1)}+\frac12\nabla^a \nabla^a \Phi\fud{a(s-2)m}{m}-m^2\Phi^{a(s)}+2\Lambda g^{aa}\Phi\fud{a(s-2)m}{m}=0\,,
\end{align}
where the mass is $m^2=-\Lambda s(s-3)$ and $\Lambda$ is the cosmological constant that we usually set to $1$. The free action is also known \cite{Fronsdal:1978rb,Buchbinder:2001bs}. The Fronsdal approach is regarded as a higher spin generalization of the metric approach to Gravity. 

The second approach is to generalize vielbein and spin-connection to higher spin fields \cite{Aragone:1979hx,Aragone:1980rk,Vasiliev:1980as}. Additional simplifications occur in three dimensions \cite{Blencowe:1988gj}. The higher spin cousins of dreibein and spin-connection are one-forms $e^{a(s-1)}\equiv e^{a_1...a_{s-1}}_\mu\, dx^\mu$ and $\omega^{a(s-1)}\equiv \omega^{a_1...a_{s-1}}_\mu\, dx^\mu$ that are symmetric and traceless in $a_1...a_{s-1}$. That the spin-connection $\omega$ looks identical to the dreibein $e$ is a genuine $3d$ effect that is well-known alredy in the case of gravity, where $\omega^a = \epsilon\fud{a}{bc}\omega^{b,c}$. The Fronsdal equations put into the first order form read
\begin{align}\label{masslesslorentz}
    \nabla e^{a(s-1)} +\epsilon\fud{a}{bc} h^b \wedge \omega^{a(s-2)c}&=0\,,&
    \nabla \omega^{a(s-1)} +\epsilon\fud{a}{bc} h^b \wedge e^{a(s-2)c}&=0\,.
\end{align}
The Fronsdal field is embedded as the totally symmetric part of the dreibein
\begin{align}
    \Phi^{a_1...a_s}&=e^{a_1...a_{s-1}}_\mm h^{\mm a_s}+\text{symmetrization}\,.
\end{align}

At this point it is convenient to switch to the spinorial language. A traceless rank-$s$ $so(2,1)$-tensor $T_{a_1...a_s}$ corresponds to a rank-$2s$ $sl_2(\mathbb{R})$-tensor $T_{\ga_1...\ga_{2s}}$. Here, $\ga,\gb,...=1,2$ are the indices of $sl_2$ or spinor indices of $so(2,1)$. The map between the $sl_2$-base and the $so(2,1)$-base is via Pauli matrices, $\sigma_m^{\ga\gb}$. After translation to the spinorial language is done we find\footnote{Our convention is that spinorial indices are raised and lowered with $\epsilon_{\ga\gb}=-\epsilon_{\gb\ga}$, $\epsilon_{12}=1$ as follows: $T^\ga=\epsilon^{\ga\gb}T_\gb$, $T_\ga=T^\gb\epsilon_{\gb\ga}$. } 
\begin{align}\label{framespinor}
    \nabla e^{\ga(2s-2)}+h\fud{\ga}{\gb}\wedge \omega^{\gb \ga(2s-3)}&=0 \,,&
    \nabla \omega^{\ga(2s-2)}+h\fud{\ga}{\gb}\wedge e^{\gb \ga(2s-3)}&=0\,.
\end{align}
One more simplification can be achieved by making the gauge algebra of pure gravity, $sl_2\oplus sl_2$, manifest. The $AdS_3$ symmetry algebra is $sl_2\oplus sl_2$ and the torsion and curvature constraints for the background $AdS_3$ dreibein $h^{\ga\gb}$ and spin-connection $\varpi^{\ga\gb}$
\begin{align}
    dh^{\ga\gb}+\varpi\fud{\ga}{\gc}\wedge h^{\gb\gc}&=0\,, & d\varpi^{\ga\gb}+\varpi\fud{\ga}{\gc}\wedge \varpi^{\gb\gc}+h\fud{\ga}{\gc}\wedge h^{\gb\gc}&=0\,,
\end{align}
can be rewritten simply as ($A_L=\varpi+e$, $A_R=\varpi-e$)
\begin{align}
    dA^{\ga\gb}_L+A_L\fud{\ga}{\gc}\wedge A^{\gb\gc}_L&=0\,, & dA^{\gad\gbd}_R+A_R\fud{\gad}{\gdd}\wedge A^{\gbd\gdd}_R&=0\,,
\end{align}
where from now on it will be useful to distinguish between the two $sl_2$ subalgebras. In particular, we reserve indices $\ga,\gb,...$ for the first $sl_2$ and indices $\gad,\gbd,...$ for the second one. As a result, instead of coupled equations \eqref{framespinor}, in the diagonal base we find two decoupled covariant constancy conditions
\begin{align}
    D_L \Omega^{\ga(2s-2)}_L&=0\,, & D_R \Omega^{\gad(2s-2)}_R&=0\,,
\end{align}
where $D_L$ and $D_R$ are the usual covariant derivatives with respect to $A_L$ and $A_R$. We note that $(D_L)^2=(D_R)^2=0$ and the gauge transformations are $\delta \Omega_{L,R}=D_{L,R}\xi_{L,R}$. 

Summarizing the dictionary, a massless spin-$s$ field can be described either by the Fronsdal field or by two connections:
\begin{align}\label{masslessdict}
    &\delta\Phi^{a(s)}=\nabla^a \xi^{a(s-1)} && \Longleftrightarrow && \delta \Omega^{\ga(2s-2)}=D_L \xi^{\ga(2s-2)}\,,\quad \delta \Omega^{\gad(2s-2)}=D_R \xi^{\gad(2s-2)}\,.
\end{align}

\paragraph{Partially-massless fields.} Partially-massless fields \cite{Deser:1983mm,Higuchi:1986wu,Deser:2001us} require non-zero cosmological constant and extend the class of massless fields. For a rank-$s$ symmetric tensor field there are $s$ (partially)-massless options parameterized by the number of derivatives in gauge transformations:
\begin{align}
    \delta\Phi^{a(s)}&=\overbrace{\nabla^a ...\nabla^a}^t \xi^{a(s-t)}-\text{traces} && t=1,...,s\,.
\end{align}
Here we assumed that the transverse-traceless gauge is imposed. The gauge-fixed equations of motion are still second-order as for massless fields,
\begin{align}
    (\square -m^2) \Phi^{a(s)}&=0\,, && m^2=-\Lambda((s-t+1)(s-t-1)-s)\,.
\end{align}
The free action is quite cumbersome due to the need for many auxiliary fields \cite{Zinoviev:2001dt}. 

Partially-massless fields admit a frame-like description \cite{Skvortsov:2006at} and the set of frame-like fields simplifies a lot in three dimensions to give \cite{Buchbinder:2012bz}
\begin{align}\label{longset}
    &e^{a(s-t)}\,, && e^{a(s-t+1)} && ... && e^{a(s-1)} \,; &&
    &\omega^{a(s-t)}\,, && \omega^{a(s-t+1)} && ... && \omega^{a(s-1)} \,.
\end{align}
The free action or equations of motion are also cumbersome since they couple the neighbouring fields together and can be found in \cite{Skvortsov:2006at,Buchbinder:2012bz}. A key observation is that the set of connections needed to describe a partially-massless field forms just two irreducible representations\footnote{Let us stress that the set in \eqref{longset} consists of the Lorentz tensors, i.e. they are representations of the diagonal $sl_2$. The second $sl_2$ mixes them together. In the $sl_2\oplus sl_2$ base we find instead just two representations as in \eqref{pmframe}. } of $sl_2\oplus sl_2$
\begin{align}\label{pmframe}
    &\delta\Phi^{a(s)}=\overbrace{\nabla^a ...\nabla^a}^t \xi^{a(s-t)} && \Longleftrightarrow && \begin{cases} \delta\Omega^{\ga(2s-t-1),\gad(t-1)}=D\xi^{\ga(2s-t-1),\gad(t-1)}\,,\\ \delta\Omega^{\ga(t-1),\gad(2s-t-1)}=D\xi^{\ga(t-1),\gad(2s-t-1)}\,.\end{cases}
\end{align}
In terms of the new variables the equations take a very simple form 
\begin{align}
    D\Omega^{\ga(2s-t-1),\gad(t-1)}&=0\,, & D\Omega^{\ga(t-1),\gad(2s-t-1)}&=0\,,
\end{align}
where $D$ is the $sl_2\oplus sl_2$ covariant derivative in this module:
\begin{align}
    D \Omega^{\ga(2j_1),\gad(2j_2)}\equiv d\Omega^{\ga(2j_1),\gad(2j_2)}+A_L\fud{\ga}{\gb} \wedge \Omega^{\ga(2j_1-1)\gb, \gad(2j_2)}+A_R\fud{\gad}{\gbd} \wedge \Omega^{\ga(2j_1),\gad(2j_2-1)\gbd}\,.
\end{align}

From the general point of view the massless case is a degenerate one since each of the two connections carries a nontrivial irreducible representation of one of the two $sl_2$ subalgebras. The degeneracy is lifted for $t>1$. Without further ado, it is clear that the actions in \cite{Skvortsov:2006at,Buchbinder:2012bz} can be rewritten as
\begin{align}\label{simplepmaction}
    S_2&= \int \Omega^{\ga(2s-t-1),\gad(t-1)} \wedge D \Omega_{\ga(2s-t-1),\gad(t-1)}-
    \Omega^{\ga(t-1),\gad(2s-t-1)} \wedge D\Omega_{\ga(t-1),\gad(2s-t-1)}
\end{align}
which also covers the massless case. Note that, as different from the massless case, we do not have any simple $e\pm\omega$ change of variable for the partially-massless case that maps the frame-like action in terms of \eqref{longset} to \eqref{simplepmaction}. 

\paragraph{Conformal fields.} The last class of fields we would like to consider are conformal or Fradkin-Tseytlin fields \cite{Deser:1983tm,Erdmenger:1997wy,Vasiliev:2009ck, Bekaert:2013zya, Beccaria:2015vaa, Kuzenko:2019ill}. Conformal fields can naturally be considered both in Minkowski and anti-de Sitter backgrounds. Free conformal fields are specified by spin $s$ and depth $t$, which is similar to the partially-massless case. Free gauge transformations read
\begin{align}\label{conforfield}
    \delta \Phi_{a_1...a_s} &= \nabla_{a_1}...\nabla_{a_t}\xi_{a_{t+1}...a_s}-\text{traces}
\end{align}
and both the field and the gauge parameter are assumed to be traceless. The equations of motion have $(2s-2t+1)$ derivatives, see \cite{Pope:1989vj,Henneaux:2015cda,Kuzenko:2016qwo,Kuzenko:2016qdw,Basile:2017mqc,Buchbinder:2018dou,Kuzenko:2019ill,Buchbinder:2019yhl}.

The frame-like description is very similar to the partially-massless case \cite{Skvortsov:2006at}. The general rule is that the frame-like field is a one-form connection that takes values in a representation of the spacetime symmetry algebra associated with the global reducibility parameters, see e.g. \cite{Barnich:2015tma}. The latter are the gauge parameters that leave the gauge field intact. For \eqref{conforfield} they are given by conformal Killing tensors. Therefore, one needs to take a one-form that, as a fiber tensor, carries an irreducible representation of the conformal algebra $so(3,2)$ corresponding to the Young diagram with rows of length $s-1$ and $s-t$: 
\begin{align}\label{conformaldiagram}
    &\Omega^{A(s-1), B(s-t)} && \Longleftrightarrow && \parbox{90pt}{\begin{picture}(90,20)\put(0,0){\RectBRow{9}{7}{$s-1$}{$s-t$}}\put(0,0){\YoungAA}\put(60,0){\YoungA}\put(80,10){\YoungA}\end{picture}}\,,
\end{align}
where $A,B,...=0,...,4$ are the indices of $so(3,2)$. Splitting $A=a,+,-$ one can decompose $\Omega$ into a number of frame-like fields that are tensors of the Lorentz algebra $so(2,1)$. The higher spin dreibein is a particular component in this decomposition: 
\begin{align}
    \Phi^{a_1...a_s}&=\Omega^{a_1...a_{s-1},+(s-t)}_\mm h^{\mm a_s}+\text{symmetrization}-\text{traces}\,,
\end{align}
which establishes a dictionary with the Fradkin-Tseytlin fields. The equations of motion are equivalent to
\begin{align}
    D\Omega^{A(s-1), B(s-t)}&=0\,,
\end{align}
where $D=d+A$, $D^2=0$ is the background covariant derivative and $A\equiv A^{A,B}_\mm dx^\mm$ is a flat connection of $so(3,2)$. We recall that the $3d$ conformal gravity can also be formulated as Chern-Simons theory for $A$, \cite{vanNieuwenhuizen:1985cx,Horne:1988jf}. Note that both $AdS_3$ and Minkowski spaces correspond to a certain $A$ such that $A^{a,+}_\mm=h^a_\mm$ is a nondegenerate dreibein. 

Summarizing, the dictionary between the metric-like and frame-like formulations in the case of conformal higher spin fields reads
\begin{align}\label{conformaldict}
    \delta \Phi_{a_1...a_s} &= \nabla_{a_1}...\nabla_{a_t}\xi_{a_{t+1}...a_s}+... && \Longleftrightarrow && \delta \Omega^{A(s-1), B(s-t)}=D\xi^{A(s-1), B(s-t)}
\end{align}
In the subsequent sections we will study interacting theories for massless, partially-massless and conformal higher spin fields. 

\section{Higher Spin Gravities in three dimension}
\label{sec:hsgravity}

The main claim of the paper is that all background independent higher spin theories in three dimension with (partially)-massless or conformal higher spin fields and without propagating matter fields have the form of the flatness condition and, provided there is an non-degenerate invariant bilinear form, the equations can be obtained from the Chern-Simons action 
\begin{align}
    S[\Omega]&=\int \Tr\left[\Omega \wedge d \Omega +\frac23 \Omega\wedge \Omega\wedge \Omega\right]\,,
\end{align}
for an appropriate higher spin extension of the anti-de Sitter $\mathfrak{g}=sl_2\oplus sl_2$ or conformal $\mathfrak{g}=so(3,2)$ algebras. By a higher spin extension of some semi-simple $\mathfrak{g}$ we mean any Lie algebra $\hs$ such that $\mathfrak{g}\subset\hs$ and the decomposition of $\hs$ into $\mathfrak{g}$-modules contains representations bigger than $\mathfrak{g}$ itself (seen as the adjoint one). Given such an algebra we can take $\hs$-valued connection $\Omega$ and write down the flatness condition. The dictionary presented in section \ref{sec:freefields} allows us to identify each $\mathfrak{g}$-submodule of $\hs$ with a particular (partially)-massless or conformal higher spin field. If $\hs$ has a non-degenerate bilinear form the equations can be obtained from the Chern-Simons action, otherwise we have equations of motion only.

This statement is highly nontrivial from the metric-like point of view. Once the equivalence between the frame-like and metric-like formulations is established it is almost a folklore that the Chern-Simons action is the unique solution of the problem. We leave the proof to section \ref{sec:equivalence} and consider below a large class of theories. The main new result here is a new class of higher spin theories with partially-massless fields. 

\subsection{Higher Spin Algebras}
\label{subsec:hsa}
Higher spin algebras seem to always originate from associative algebras.\footnote{We are not aware of a any example of a higher spin algebra that does not come from an associative algebra via the construction given below.} There is a large class of associative algebras that contain a given Lie $\mathfrak{g}$ algebra as a Lie subalgebra. The class is parameterized by various irreducible modules of $\mathfrak{g}$. Given a $\mathfrak{g}$-module $V$ we can simply take $\mathrm{End}(V)= V\otimes V^*$ as an associative algebra. The same algebra can be understood as a quotient of the universal enveloping algebra of $\mathfrak{g}$ modulo the two-sided ideal $\mathrm{Ann}(V)$ that annihilates $V$ (the annihilator):
\begin{align}
   \text{associative}&: &\hs(V)&= \mathrm{End}(V)= V\otimes V^*= U(\mathfrak{g})/\mathrm{Ann}(V)\,.
\end{align}
If $V$ is infinite-dimensional, some care is needed in working with the, otherwise equivalent, definitions above. 

We would like to highlight several features of $\hs(V)$. Firstly, the construction gives $\hs(V)$ as an associative algebra. Since we are interested in the algebras relevant for the Chern-Simons formulation, only its induced Lie structure, which is obtained via commutators, will be needed. As a Lie algebra we have
\begin{align}\label{Liedecomposition}
    \text{Lie}&: &\hs(V)&= gl(V)=sl(V)\oplus u(1)\,.
\end{align}
The $u(1)$-factors lead to abelian Chern-Simons fields that decouple. Secondly, the above class of higher spin algebras admits a simple generalization where the $u(1)$ field turns into a non-abelian one \cite{Gwak:2015vfb}. More precisely, one can tensor $\hs(V)$ with (usually semi-simple and usually finite-dimensional) associative algebras, i.e. matrix algebras. Then, one can take truncate the resulting Lie algebra with the help of some (anti)-automorphisms and impose certain reality conditions, see e.g. \cite{Konstein:1989ij}. This way, for example, one can get $so(V)$ and $sp(V)$ truncations of $\hs(V)$. By construction $\hs(V)$ is equipped with a non-degenerate invariant bilinear form.

An interesting feature of the higher spin theories with matrix algebra extensions is that the spin-two sector that corresponds to gravity admits new $(A)dS$ solutions with different cosmological constants. Around these solutions, the spectrum of the theory restructures itself \cite{Gwak:2015vfb,Gwak:2015jdo} combining massless fields into partially-massless ones.

Annihilator $\mathrm{Ann}(V)$ is also an interesting algebra, which is usually thrown away. It is an associative algebra by construction, which can be decomposed into irreducible finite-dimensional $\mathfrak{g}$-modules. Therefore, $\mathrm{Ann}(V)$ gives a class of higher spin algebras that do not contain $\mathfrak{g}$ as a subalgebra.\footnote{Except for the trivial case when $V$ is one-dimensional.} However, $\mathrm{Ann}(V)$ is quite big and is not multiplicity free. Indeed, for any irreducible $V$ annihilator $\mathrm{Ann}(V)$ contains generators of the form $I_i(\lambda_i)=(C_i-\lambda_i)$ where $i$ runs over all independent Casimir operators $C_i$ and $\lambda_i$ are values thereof on $V$. To reduce the multiplicity we can define a family of algebras
\begin{align}
    A_{\{\lambda\}}&= U(\mathfrak{g})/ I_{\{\lambda\}}
\end{align}
where $I_{\{\lambda\}}$ is a two-sided ideal generated by all $I_i(\lambda_i)$. At special values of $\lambda_i$ that correspond to, say, finite-dimensional module $V$, $A_{\{\lambda\}}$ develops a two-sided ideal $J_V$ such that the quotient $A_{\{\lambda\}}/J_V$ coincides with earlier defined $\hs(V)$. The ideal $J_V$ is an analog of $\mathrm{Ann}(V)$ but with the multiplicity considerably reduced. Sometimes, see below, $J_V$ is multiplicity free.\footnote{For any $\lambda_i$ we can think of the generalized Verma module $V$ that makes Casimir operators $C_i$ equal $\lambda_i$.}

If nontrivial, algebra $J_V$ leads to a class of theories that contain higher spin fields, but do not have the graviton since $J_V$ does not contain $\mathfrak{g}$. In our cases $\mathfrak{g}$ can be $sl_2\oplus sl_2$ or $so(3,2)$. Another interesting feature is that we have interacting higher spin theories that are background dependent since we cannot absorb the AdS background into a dynamical spin-two field. The equations read
\begin{align}\label{backgrounddep}
    D \Omega +\tfrac12[\Omega,\Omega]&=0\,, && D^2=0\,, && D=d+\Omega_0\,,
\end{align}
where $\Omega_0$ is the background flat connection of $\mathfrak{g}$ (we still have that $J_V$ is not only an algebra, but it is also a $\mathfrak{g}$ module). Nevertheless, we can extend $J_V$ with $\mathfrak{g}$ into a new Lie algebra $\mathfrak{f}=\mathfrak{g} \rtimes J_V$. This allows us to add the graviton into the theory, but there is no backreaction from higher spin fields to the gravitational stress-tensor. Therefore, the newly added spin-two field does not behave like a graviton.

In case we have a non-degenerate invariant bilinear form, the action is still of the Chern-Simons type
\begin{align}
    S[\Omega,D]&=\int \Tr\left[\Omega \wedge D \Omega +\frac23 \Omega\wedge \Omega\wedge \Omega\right]\,,
\end{align}
but with $d$ replaced with the background covariant derivative $D=d+\Omega_0$. We cannot absorb $\Omega_0$ into $\Omega$ since $\mathfrak{g}$ acts on $\Omega$, but it is not a subalgebra. There is a similar phenomenon in $d>3$ for Type-B,C theories \cite{Gunaydin:2016amv,Giombi:2016pvg}.

After the general comments about higher spin algebras, let us briefly discuss the two known cases: purely massless and conformal higher spin theories.  

\paragraph{Massless higher spin algebras.} There are many non-semisimple (higher spin) algebras, but a rich enough class of theories is obtained by taking any of the classical Lie algebras $su_N$, $so_N$ and $sp_N$ that can be understood as Lie subalgebras of $\hs(V)=\mathrm{End}(V)$, \cite{Blencowe:1988gj,Bergshoeff:1989ns,Campoleoni:2010zq,Henneaux:2010xg}, where $V$ is an irreducible $sl_2$-module of dimension $N$ and the $sl_2$ subalgebra corresponds to the principal embedding into $su_N$, $so_N$ or $sp_N$. Then, the action is the difference of two Chern-Simons actions for $\hs$, i.e. is a particular version of $\hs\oplus \hs$ Chern-Simons theory. The spectrum of massless (higher spin) fields can be read off from the decomposition of $\hs$ into $sl_2$ modules $V_j$, $\mathrm{dim}\, V_j=2j+1$, according to \eqref{masslessdict}:
\begin{align}
    & V_j && \Longleftrightarrow && \text{spin}=(j-1)
\end{align}
There is also a one-parameter family of associative algebras $hs(\lambda)$.\footnote{It was first defined in \cite{Feigin} and dubbed $gl_\lambda$ because it interpolates between all $gl_n$, $n=1,2,3,...$.} Using the conventions introduced above, $hs(\lambda)$ is defined as a quotient
    \begin{align}
        hs(\lambda)&=U(sl_2)/I_\lambda\,, && I_\lambda=U(sl_2)[\boldsymbol{C}_2+(\lambda^2-1)]\,.
    \end{align}
For generic $\lambda$ the algebra is infinite-dimensional and decomposes into $V_0\oplus V_1\oplus V_2 \oplus ...$. The singlet $V_0$ corresponds to $u(1)$, c.f. \eqref{Liedecomposition}, and can be removed after passing to the Lie algebra. An interesting property of $hs(\lambda)$ is that for $\lambda \in \mathbb{Z}$ it develops a two-sided ideal $J_\lambda$ such that the quotient is $gl_ {|\lambda|}$. Note that $gl_{|\lambda|}$ decomposes as 
\begin{align}\label{masslesscase}
   gl_\lambda&= V_0\oplus V_1\oplus...\oplus V_{\lambda-1}
\end{align}
with respect to the principal $sl_2$ embedding. Therefore, the ideal $J_\lambda$ decomposes as
\begin{align}
    J_\lambda&= V_\lambda \oplus V_{\lambda+1}\oplus ...
\end{align}
This gives an example of a higher spin algebra that does not contain the gravitational subalgebra, $sl_2$ in this case. Therefore, the resulting higher spin theory is background dependent, i.e. of the form \eqref{backgrounddep}. It is unclear if a non-degenerate invariant bilinear form exists (in principle, it can be obtained by dropping the leading zero in \cite{Joung:2014qya}). 

\paragraph{Conformal higher spin algebras.} The construction above can be applied to the conformal algebra $\mathfrak{g}=so(3,2)$, as was done in \cite{Grigoriev:2019xmp}. Without going into too many detail, a large class of finite-dimensional conformal higher spin algebras can be constructed by taking $V$ to be any finite-dimensional irreducible representation of $so(3,2)$ (or $so(4,1)$, the signature being irrelevant here). For example, taking $V$ to be the spinorial representation $\bullet_{1/2}$ we get
\begin{align}
    \hs(\bullet_{\tfrac12})&=\YoungpAA \oplus \YoungpA\oplus \bullet \,,
\end{align}
which was studied in \cite{Pope:1989vj}. For the vector representation we find
\begin{align}
    \hs(\YoungpA)&=\YoungpA\otimes \YoungpA= \YoungpAA \oplus \YoungpB\oplus \bullet \,,
\end{align}
and for the rank-two symmetric representation:
\begin{align}
    gl(\YoungpB)&=\YoungpB\otimes \YoungpB= \YoungpBB \oplus \YoungpCA\oplus \YoungpD\oplus \YoungpB \oplus \YoungpAA \oplus \bullet\,.
\end{align}
$\bullet$ corresponds to $u(1)$ that can be decoupled. The spectrum of conformal fields can be read off with the help of dictionary \eqref{conformaldict}. 
These algebras have also an interpretation as partially-massless higher spin algebras in $(A)dS_4$ \cite{Joung:2015jza}.

\subsection{New (Partially)-massless Higher Spin Gravities}
We would like to construct a class of higher spin theories that contain partially-massless fields. The crucial step is just to look at the dictionary \eqref{pmframe}
\begin{align}\label{pmframe2}
    &\delta\Phi^{a(s)}=\overbrace{\nabla^a ...\nabla^a}^t \xi^{a(s-t)} && \Longleftrightarrow && \Omega^{\ga(2s-t-1),\gad(t-1)}\,,\quad \Omega^{\ga(t-1),\gad(2s-t-1)}\,.
\end{align}
In general, the two connections are charged with respect to both $sl_2$, which is elusive for the purely massless case, $t=1$. Since $\mathfrak{g}=sl_2\oplus sl_2$, irreducible representations $V$ of $\mathfrak{g}$ are parameterized by two irreducible representations of $sl_2$. If the modules are finite-dimensional we have $V=V_{j_1}\otimes V_{j_2}$ and the general construction of $\hs(V)$ still works: 
\begin{align}\label{spectrumpm}
    \hs(V)&= (V_{j_1}\otimes V_{j_1}) \otimes (V_{j_2}\otimes V_{j_2})=\bigoplus_{k_1=0,1,...,2j_1}\bigoplus_{k_2=0,1,...,2j_2} V_{k_1}\otimes V_{k_2}\,.
\end{align}
The very first component in the sum $k_1=k_2=0$ corresponds to the $u(1)$-factor that decouples. Two terms with $(k_1,k_2)$ equal $(1,0)$ and $(0,1)$ give embedding of $sl_2\oplus sl_2$. The rest corresponds to massless fields, which occur for $k_1 k_2=0$, and to partially-massless fields for $k_1k_2\neq0$. 

It is worth noting at this point that (partially)-massless fields are described by conjugate pairs of $sl_2\oplus sl_2$ modules. This requirement can always be satisfied by taking $\hs(V)\oplus \hs(V^T)$ where $V^T=V_{j_2}\otimes V_{j_1}$. The action is the difference of two Chern-Simons actions for $hs(V)$ and $\hs(V^T)$. For $j_1=j_2$ algebra $\hs(V)$ is self-conjugate, i.e. contains conjugate pairs.

The massless case corresponds to $j_1=0$ or $j_2=0$. For example, for $j_1=j$ and $j_2=0$ we find exactly \eqref{masslesscase}
\begin{align}
    \hs(V)&= (V_{j}\otimes V_{j}) \otimes (V_{0}\otimes V_{0})=\bigoplus_{k=0,1,...,2j} V_{k}\otimes V_{0}= gl_{2j+1}\otimes gl_1\,.
\end{align}
The second factor is trivial and we get $\hs(V)=gl_{n}$, $n=2j+1$. Upon excluding the trivial $u(1)$ we find $sl_{n}$ with the principal embedding of $sl_2$.

It is instructive to see how the construction of the partially-massless higher spin algebras above can be explained from those in generic dimensions \cite{Boulanger:2011se,Bekaert:2013zya, Alkalaev:2014nsa, Joung:2015jza}. In $d>3$ there is a one-parameter family of such algebras defined as a quotient of $U(so(d,2))$ by a certain two-sided ideal. In $d=3$, however, due to the degeneracy caused by isomorphism $so(2,2)\sim sl_2\oplus sl_2$, there is a two-parameter family of algebras $hs(\lambda_1)\otimes hs(\lambda_2)$. Then, \eqref{spectrumpm} corresponds to $\lambda_{1,2}$ being the values of the Casimir operators on $V_{j_{1,2}}$.

\subsection{Comments on the metric-like formulation}
Going from the Chern-Simons formulation to the metric-like one is not impossible, but is very difficult in practice, see e.g. \cite{Fredenhagen:2014oua}. Several seemingly nontrivial features of the metric-like formulation get a very simple interpretation in the Chern-Simons one. Consider massless fields in $AdS_3$, for example. Schematically, the equations look as follows
\begin{align}\label{perturb}
    \nabla e + h \wedge\omega &= -\omega \wedge e\,,&  \nabla\omega+\omega\wedge\omega+e\wedge e&=0\,,
\end{align}
where $e$ is a (higher spin) dreibein, $\omega$ is a (higher spin) spin-connection, $h$ is an $AdS_3$ dreibein. Both $e$ and $\omega$ contain a number of higher spin fields, in accordance with a given higher spin algebra. The first equation is a constraint to be solved for $\omega$ order by order. The second equation is the dynamical equation for the Fronsdal fields.

One starts with a free field $e_1$ (first order) that is equivalent to a collection of Fronsdal fields $\Phi_1$. We solve for $\omega_1$ in terms of $\nabla \Phi_1$. At the next order $e_2$ is expressed in terms of $\Phi_2$ and $\omega_2$ is solved as $\nabla\Phi_2 +\Phi_1 \nabla \Phi_1$ and so on. The nonlinearities grow, but the spin-connection is always expressed in terms of the first order derivatives of the Fronsdal fields. 

It is convenient to use $sl_2$ spin $j$ instead of the spin $s$ (the rank of the Fronsdal tensor), the two being related by $j=(s-1)$. As is clear from \eqref{perturb} and from the Chern-Simons action, the vertices are constrained by the $sl_2$ tensor product rules: we cannot possibly form a singlet unless there exists a triangle with edges of lengths $j_{1,2,3}$. The same rules apply when solving for $\omega$ at higher orders: $\omega\wedge e$ can contribute $\Phi\nabla\Phi$ to $\omega$ only if a triangle can be formed. As a result, the simple cubic Chern-Simons interaction generates an infinite tower of interaction vertices in the Fronsdal formulation subject to certain selection rules. At any given order $n$, only those Fronsdal fields can form a vertex for which $V_{j_1}\otimes ...\otimes V_{j_n}$ contains the singlet $V_0$. This gives exactly the polygonal constraints discovered in \cite{Fredenhagen:2018guf}. 

The latter considerations imply the following constraints for the CFT correlation functions of higher spin currents $J_{i_1...i_s}$ that are dual to Fronsdal fields $\Phi_{a_1...a_s}$. Only those correlation functions of $J_s$ may not vanish for which $V_{j_1}\otimes ...\otimes V_{j_n}$ contains the singlet representation $V_0$ \cite{Fredenhagen:2018guf}. Note, that if some of the spins are equal and represent the same fields/operators then we find more constraints as some of the tensor products need to be projected onto the (anti)-symmetric parts thereof. 

There is one more important consequence of the fact that all matter-free higher spin gravities are of the Chern-Simons type: we have only two independent types of cubic vertices. Indeed, any massless or partially-massless theory is based on $\hs\oplus \hs$ for some $\hs\supset sl_2$. Rewriting the action in terms of dreibein $e$ and spin-connection $\omega$ instead of $\Omega_L$ and $\Omega_R$ we see, schematically, the following two cubic vertices\footnote{See \cite{Boulanger:2000ni,Boulanger:2005br} for an earlier discussion of interactions in $3d$.}
\begin{align}
    S_3^{o,e}=&\int \omega\wedge \omega\wedge \omega +e\wedge e\wedge \omega\,,  &&S_3^{e,o}=\int e\wedge \omega\wedge \omega+e\wedge e\wedge e\,.
\end{align}
In the pure gravity case the second one corresponds to the Einstein-Hilbert action with the cosmological constant and is even. The first one is odd (if we define parity by the behaviour under $\omega \rightarrow -\omega$). As is discussed above, in the Fronsdal formulation both types of vertices generate an infinite number of metric-like vertices, cubic, quartic and so on. It is also clear that $S_3^{o,e}$ leads to vertices with three derivatives followed by a one-derivative term, while $S_3^{e,o}$ leads to vertices with two derivatives followed by a zero-derivative term. This is in accordance with the classification of \cite{Mkrtchyan:2017ixk, Kessel:2018ugi} provided the definition of parity is related to the number of $\epsilon$-tensors in the metric-like formulation. 

Therefore, we obtain a highly nontrivial result from the metric-like point of view: (1) there are only two independent cubic vertices for any given three spins $j_{1,2,3}$ that can form a triangle; (2) there are no independent higher order vertices, while the cubic ones entail higher order vertices such that one can draw a polygon with edges of length $j_i$ \cite{Fredenhagen:2018guf}.  

The same statements are true for partially-massless fields with the obvious replacement of $j=s-1$ with $j^L=s-(t+1)/2, j^R=(t-1)/2$. There are two independent cubic vertices for any given three spins $s_{1,2,3}$ and depths $t_{1,2,3}$ iff the tensor products $V_{j_1^L}\otimes V_{j_2^L}\otimes V_{j_3^L}$ $V_{j_1^R}\otimes V_{j_2^R}\otimes V_{j_3^R}$ contain the singlet. There are no independent higher order vertices. This gives a classification of vertices involving (partially)-massless fields.

Analogously, the only independent vertices of conformal higher spin fields are cubic ones and they are in one-to-one with the singlets in the tensor product of $so(3,2)$-modules described in \eqref{conformaldiagram}. Note, that the tensor product of two $so(3,2)$-modules is not multiplicity free in general. Therefore, as different from the (partially)-massless case, there can be several independent cubic vertices of three given conformal fields.

\section{Bootstrapping  \texorpdfstring{$\boldsymbol{3d}$}{3d} higher spin theories}
\label{sec:equivalence}
Using Fronsdal fields as an example we now discuss in some details the relation between metric-like and frame-like formulation within the BV-BRST approach. In particular, we spell out explicitly the relation between cubic vertices in these formulations in generic dimension. In the case of theory without local degrees of freedom and without nontrivial reducibility relations among gauge transformations the structure of the theory can be explicitly described at the level of equations of motion. In 3d and
under usual assumptions (which hold for (partially)-massless and conformal fields)
the system takes the form of a Chern-Simons theory.

\subsection{BV-BRST formulation of Fronsdal fields}
\label{BRST-Fronsdal}

The conventional approach to constructing BV-BRST formulation of Fronsdal fields on Minkowski space is to start with Fronsdal Lagrangian or equations of motion and build the BV-BRST formulation following the standard prescription. However an equivalent and concise BV-BRST description can be constructed starting with the partially gauge-fixed formulation, where
\begin{align}
\label{const}
 \Box \Phi&=\d_x\cdot \d_p \Phi=\d_p\cdot \d_p\Phi=0\,,&&\delta \Phi =p\cdot \d_x \Xi\,.
\end{align} 
Here we use generating functions $\Phi(x|p)$ and $\Xi(x|p)$ for fields and gauge parameters (Taylor coefficients in $p^a$ encode Fronsdal fields). Gauge parameters $\Xi$ are also subject to the same equations as $\Phi$. This gauge is known as the (on-shell) transverse-traceless gauge. Nevertheless, one can show that starting from~\eqref{const} one can, in fact, reconstruct a fully gauge invariant formulation using the parent formalism~\cite{Barnich:2004cr,Barnich:2010sw,Grigoriev:2012xg} so that there is no loss of generality.

Replacing $x^a$ with formal variables $y^a$ (i.e. generating functions are formal power series in $y$) and treating component fields of $\Xi$ as ghost fields, equations~\eqref{const} determine the BRST jet-space for the system. Note that strictly speaking this is not a jet-bundle but rather its subbundle because  coefficients of $\Phi$ and $\Xi$ are subject to differential (in $y^a)$ constraints similar to \eqref{const}. 

The jet-space is coordinatized by components of $\Phi$ and $\Xi$ as well as by spacetime coordinates $x^a$ and their differentials $\theta^a=dx^a$ which we treat as Grassmann-odd coordinate of ghost degree $1$. There are two differentials (odd nilpotent vector fields of ghost degree $1$) defined on the jet-space functions. The first is the BRST differential encoding gauge transformations and the second one is the horizontal differential encoding the equations of motion. The BRST differential is given by:
\begin{align}
\gamma \Phi&=(p\cdot \d_y) \Xi\,, && \gamma \Xi=0\,.
\end{align}
For instance for the spin-two field $\phi^{ab}$ contained in $\Phi$ one gets
\begin{equation}
\gamma \phi^{ab; c_1...c_k}=\xi^{a;b c_1...c_k}+\xi^{b;ac_1...c_k}\,,
\end{equation} 
where the convention is to put spin indices first and those associate to $y$-variables (i.e. derivatives) after the separator. The horizontal differential has the form $\Dh=\theta^a D_a$ where the action of the total derivative operator $D_a$ on coordinates on the jet-space is defined via:
\begin{align}
&D_a\Phi=\dl{y^a}\Phi\,, &&  D_a\Xi=\dl{y^a}\Xi\,, && D_a x^b=\delta^b_a\,. 
\end{align}
Functions on the above jet-space form a particularly useful version of the BRST complex for Fronsdal fields (of course, it is not unique and is defined up to equivalence). It is convenient to introduce total differential $Q_0=\Dh+\gamma$ which carries one unit of ghost degree. The ghost degree is determined by prescribing $\gh{\Xi}=1$ and $\gh{\theta^a}=1$.

The jet-space equipped with the ghost degree and total differential $Q_0$ encodes all the information about the gauge theory. In particular, equations of motion and gauge symmetries can be read-off~\cite{Barnich:2010sw} from $Q_0$. More precisely, if $\Psi^{A_k}$ collectively denote all the ghost degree $k$ coordinates on the jet-space save for spacetime coordinates $x^a$ and their differentials $\theta^a$, then one promotes each $\Psi^{A_k}$ to a field $\Phi^{A_k}(x,\theta)$ of homogeneity degree $k$ in $\theta$, i.e. it can be seen as a spacetime $k$-form with $k=\gh{\Psi^{A_k}}$. We also assume that $\Phi^{A_k}=0$ for $k<0$. Note though that for the higher spin system under considerations negative degree $\Psi$ are not present anyway. Then one subjects $\Phi^{A_k}$ to the following equations~\cite{Barnich:2010sw,Grigoriev:2019ojp}: 
\begin{equation}
\label{Q-map-eom}
 d_X(\Phi^{A_k}(x,\theta))=\left(Q_0 \Psi^{A_k}\right)\big|_{\Psi^{B_l}=\Phi^{B_l}(x,\theta)} \,.
\end{equation} 
Here $d_X$ is the exterior differential. Note that for $k=-1$ the LHS is trivial while for $k<-1$ both LHS and RHS vanish identically. In a similar way one defines gauge transformations. 
The above system can be seen as a far-going generalization of the AKSZ-type sigma model~\cite{Alexandrov:1995kv} and in fact can be inferred from AKSZ equations of motion if one starts with the parameterized system to begin with, see~\cite{Grigoriev:2012xg} for more details. 

In contrast to the conventional BV-BRST approach to local gauge theories, which operates in terms of jet-bundles, in the present context we employ more general underlying spaces (roughly speaking those with differential constraints on fields and ghosts) and more flexible notion of equivalence which does not respect the decomposition of $Q_0$ into space-time part and field-space parts. This allows for a very concise formulations of the theory. Remarkably, one can always reconstruct a usual field theoretical formulation of the system through e.g.~\eqref{Q-map-eom} just in terms of the total differential and in terms of fields valued in the underlying (possibly constrained) jet-space. This approach to general gauge theories was originally developed~\cite{Barnich:2004cr,Barnich:2010sw,Grigoriev:2010ic,Grigoriev:2012xg} under the name of parent formulation. Its more invariant and geometrical version was proposed recently in~\cite{Grigoriev:2019ojp},
where it was also explicitly related to the invariant approach to PDEs~\cite{Vinogradov1981} (for a review see e.g.~\cite{Anderson1991,Krasilshchik:2010sst}). In particular, the BRST complex with total differential $Q_0$ can be seen as a BRST extension of the infinitely prolonged PDE. It is also worth mentioning close relation to the unfolded formalism~\cite{Vasiliev:1988xc,Vasiliev:2001wa,Vasiliev:2005zu} developed in the context of HS theory.

Equations~\eqref{Q-map-eom} have a simple geometrical interpretation~\cite{Grigoriev:2019ojp}: fields are components of a section 
$\sigma:T[1]X \to \text``jet-space''$ (jet-space is naturally a bundle over $T[1]X$, i.e. the spacetime $X$ extended by $\theta^a$), i.e. $\Phi^{A_k}(x,\theta)=\sigma^*(\Psi^{A_k})$, where $\sigma^*$ is a pullback map induced by $\sigma$,
while the equations of motion~\eqref{Q-map-eom} say that $\sigma$ is a $Q$-map (i.e. $d_X \circ \sigma^*= \sigma^*\circ Q_0$, or in other words $\sigma^*$ is a map of the respective homological complexes). Gauge transformations correspond to trivial deformations of $\sigma$, i.e. those of the form $\delta_\epsilon \sigma^*=d_X\circ \epsilon^* + \epsilon^*\circ Q_0$ for some map $\epsilon^*$ of degree $-1$, which encodes  gauge parameters.

In the case at hand the above procedure amounts to promoting $\Phi$ to a spacetime field $\Phi(x|y,p)$ and $\Xi$ to a one-form field $\Xi(x,\theta)=\theta^b\, \Xi_b(x|y,p)$. In these terms the equations 
of motion take the form~\cite{Barnich:2004cr}:
\begin{equation}
 (d_X-\theta^a\dl{y^a})\Phi=p\cdot \d_y\, \Xi\,, \qquad  (d_X-\theta^a\dl{y^a})\Xi=0\,.
\end{equation} 
Note that it goes without saying that the $y$-space version of \eqref{const} is imposed on $\Phi$, $\Xi$. Further details can be found e.g. in section 5 of~\cite{Grigoriev:2012xg} and references therein.

\subsection{Minimal model}
When studying local BRST cohomology it can be very convenient to work with the "minimal" version of the BRST complex. Practically, this can be obtained by eliminating the maximal amount of contractible pairs for the total BRST differential $Q_0$. This approach was extensively used in~\cite{Barnich:1995ap,Brandt:1996mh,Brandt:1997iu,Brandt:2001tg}, though the idea of using the total differential and somewhat implicit version of the minimal model for usual gauge theories was already in~\cite{Stora:1983ct,Manes:1985df}.

If by elimination of contractible pairs the underlying jet-space remains the bundle over the spacetime manifold extended by basis differentials $\theta^a$ then such elimination is an equivalence not only of homological complexes  but also of local gauge field theories. In particular, one can reconstruct an equivalent formulation of the theory in terms of the reduced complex. This gives a powerful tool to construct new equivalent formulations of a given gauge system. 

In the case at hand the minimal formulation is constructed by eliminating the maximal amount of contractible pairs for $Q_0$. Contractible pairs are easily identified as originating from those for the BRST differential $\gamma$. Since $\gamma$ is determined  by the operator $p\cdot \dl{y}$, coordinates that are not contractible pairs are associated to the kernel and cokernel of this operator. These coordinates are encoded in the generating functions satisfying $p\cdot \dl{y}\Xi=0$ and $y\cdot \dl{p}\Phi=0$, so that the component fields are precisely the familiar tensors \cite{Lopatin:1987hz} associated to the two-row Young tableaux. 

Denoting by $\bar \Xi$ and $\bar\Phi$ generating functions  for the remaining fields in ghost degree $1$ and $0$, respectively, the reduced differential is given by \cite{Barnich:2004cr}
\begin{align}\label{minimaleqs}
 \bar Q_0\bar\Phi&=\Pi (\theta \cdot\dl{y})\bar\Phi\,,&& \qquad \bar Q_0\bar\Xi=(\theta \cdot\dl{y})\bar\Xi+\mu(\bar\Phi)\,, &&\qquad Q_0 x^a=\theta^a\,,
\end{align} 
where $\Pi$ denotes the projector onto the kernel of $y\cdot\dl{p}$ and the last term is linear in $\bar\Phi$ and quadratic in $\theta^a$. Note that in the minimal formulation $\bar Q_0$ does not respect the form degree and, hence, it cannot be represented as a sum of the spacetime and of the field parts. In particular, interaction as well as other physical objects are described by $Q_0$-cohomology.

The equations of motion~\eqref{Q-map-eom} determined by the above $\bar Q_0$  is nothing but the equations of the unfolded formulation of Fronsdal fields, which was originally arrived at from different perspective long before, see~\cite{Vasiliev:1988xc,Vasiliev:2001wa,Vasiliev:2005zu} and references therein. Note that although we started with the metric-like formulation of the Fronsdal system what we have arrived at by resorting to the BRST description followed by a reduction to its minimal model is the unfolded formulation, which is (an extension of) the frame-like one. As we are going to see the same happens at the level of interaction vertices. The interactions in terms of the minimal BRST formulation naturally reproduce frame-like vertices. More precisely, frame-like vertices correspond to $\bar Q_0$-cohomology of degree $d$. Given a $\bar Q_0$-cocycle $V$ the explicit form of the vertex can be written as~$\int_{T[1]X} \sigma^*(V)$. In this work we refrain from discussing explicit realization of frame-like vertices in $d>3$ and postpone the discussion of 3d frame-like vertices for Section~\bref{sec:metric2frame}.

The above discussion applies to Fronsdal fields in any spacetime dimension $d$. In 3d $\bar\Phi$ vanish for spin grater than $1$. In what follows we assume that spin-$0$ is not present while we set $\bar\Phi=0$ by hands for spin-one. This of course amounts to considering the topological spin-one field rather than Maxwell spin-one field.

Under these assumptions the minimal model for the BRST complex takes a rather concise form:
\begin{align}\label{mincomplex}
 \bar Q_0x^a&=\theta^a\,,&& \bar Q_0\theta^a=0\,,&&  \bar Q_0 \bar\Xi=\theta \cdot \dl{y} \bar\Xi
\end{align} 
Let us recall that generating function $\bar\Xi(y,p)$ is subject to $p\cdot\dl{y}\bar\Xi=0$ along with $\dl{p}\cdot\dl{p}\bar\Xi=0$, giving the irreducibility conditions on the coefficients. Note that the equations of motion~\eqref{Q-map-eom} with $Q_0$ replaced with the above $\bar Q_0$  is precisely the frame-like  \eqref{masslesslorentz} or \eqref{framespinor} 
equations in flat space if we identify 1-form fields entering $\bar\Xi$ as $e$ and $\omega$.\footnote{In \eqref{mincomplex} we do not take an advantage of dualizing the spin-connection $\omega^{a_1...a_{s-1},b}$ into $\omega^{a_1...a_{s-1}}$ as in \eqref{masslesslorentz}. See also below for the $sl_2$ realization of the same modules. }

In terms of the minimal model it is easy to switch from flat space to the constant curvature space (for definiteness AdS space). The only difference is that $\dl{y^a}$ in $\theta\cdot \dl{y}$ gets modified into a certain linear operator $\omega_a$ acting on the linear space $\algA$ associated to $\bar\Xi$ (in the case at hand this is the space of polynomials in $y^a,p^a$ annihilated by 
$p\cdot\dl{y}$ and $\dl{p}\cdot\dl{p}$; of course the same space is more conveniently described in terms of $sl_2$ tensors). In these terms coefficients $A^I$ of $\bar\Xi$
can be seen as coordinates on $\algA[1]$, i.e. a supermanifold associated to $\algA$ and whose coordinates are odd and of ghost degree $1$.  Then 
\begin{align}
\label{minicomplex2}
Q_0x^a&=\theta^a\,,&& Q_0\theta^a=0\,,&& Q_0A^I=\theta^m\omega_m{}^I{}_J A^J\,.
\end{align}
Of course, $\omega_m{}^I{}_J$ have a meaning of the coefficients of a flat connection of AdS algebra (in the representation $\algA$). A systematic derivation of the BRST description of Fronsdal fields in AdS space can be found in~\cite{Barnich:2006pc}.

Furthermore, the minimal BRST complex for (partially)-massless or conformal fields in 3d is also of the form~\eqref{minicomplex2}. The only difference is that 
the module $\algA$ has to be replaces with the respective module of the global symmetry algebra and $\omega_m{}^I{}_J$ 
with the coefficients of a flat connection of the symmetry algebra describing the background geometry. Both $\algA$ and 
$\omega_m{}^I{}_J$ can be read-off from the frame like description reviewed in Section~\bref{sec:freefields}. That the resulting minimal BRST complex in this case is equivalent to the conventional BRST complex in terms of metric-like fields can be immediately checked by e.g. obtaining the minimal form of BRST complexes for partially-massless and conformal fields that can be taken from e.g.~\cite{Alkalaev:2009vm,Alkalaev:2011zv} and \cite{Bekaert:2013zya} respectively.

As we reviewed in Section \bref{sec:freefields}, in the case of fields on $AdS_{3}$, it is more useful to describe $\algA$ in terms of $sl_2$ tensors. More precisely, a spin-$s$ field gives rise to $\algA$, which runs over a direct sum of two $sl_2$-modules $V_{s-1}$, i.e. $A^{\ga(2s-2)}$, $A^{\gad(2s-2)}$. In general, for (partially)-massless fields we have $\algA$ given by $V_{2s-t-1} \otimes V_{t-1} \oplus  V_{t-1} \otimes V_{2s-t-1}$ of $sl_2\oplus sl_2$.\footnote{Here we adopt different notations compared to \cite{Gwak:2015jdo}. Here, the massless case corresponds to $t=1$, while there it was $t=0$. The finite-dimensional modules $V_{t}$ here correspond to $R_{\tfrac{t+1}2}$ there.} At this point we do not have to make any assumptions about the spectrum of fields that $I$ runs over, i.e. about $\algA$. Also, even though we are primarily concerned with the (partially)-massless fields, conformal fields are also covered by taking $I$ to run over $so(3,2)$-modules described in Section \bref{sec:freefields}. In the latter case $\omega$ is an $so(3,2)$ flat connection.

More generally, the structure of the linear BRST complex~\eqref{minicomplex2} is unchanged even if we consider a general linear gauge system in generic dimension without local degrees of freedom (so that the only degree-zero coordinates are $x^a$) and no nontrivial reducibility identities between gauge generators (so that there are no coordinates in of degree $2$ and higher). It follows, the discussion of possible nonlinear completions given in the following sections  fully applies to generic theories of this sort. The only difference is that $\algA$ and the flat connection describing the background are different. 

\subsection{Deformation theory and interactions} 
Suppose there is an interacting theory that have total differential $Q_0$ as a linearization. Expanding $Q$ in powers of fields one gets $\commut{Q_0}{Q_1}=0$. At the same time, trivial deformations correspond to $Q_1=\commut{Q_0}{T_1}=0$ for some $T_1$ with $\gh{T_1}=0$, i.e. interactions are controlled by cohomology of $Q_0$ in vector fields of ghost degree $1$. This is a nonlagrangian version~\cite{Barnich:2004cr} of the standard BV-BRST approach to consistent deformations of Lagrangian gauge theories~\cite{Barnich:1993vg}.

If the theory in question admits Lagrangian formulation (the one we are talking about does) the interactions are parameterized by the $Q$-cohomology in the space of local functions of ghost degree $d$ (spacetime dimension)~\cite{Barnich:1993vg}. 

It is important to stress that $Q$-cohomology is invariant under elimination of contractible pairs. That is why one can use any formulation, not necessarily Lagrangian (e.g. minimal) to compute the cohomology. 

It turns out that in the case at hand (conformal or (partially)-massless fields in 3d or generic theory without local degrees of freedom and nontrivial reducibility identities) the problem of cubic vertices is substantially simplified because in the minimal formulation the only coordinates of vanishing ghost degree are spacetime coordinates $x^a$. More precisely, the BRST complex is given by the algebra of functions in degree 1 coordinates $A^I$, $\theta^a$ and degree $0$ coordinates $x^a$. 

If one in addition insists on the translation invariance, cocycles cannot depend on $x^a$ so that it is enough to analyze the cohomology of the total differential $Q_0$ in the Grassmann algebra generated by $A^I$, $\theta^a$. In particular, those of degree $3$ (and hence cubic in the coordinates) correspond to nontrivial vertices. Despite the fact that quartic vertices cannot appear there is a consistency condition at the next order ensuring that the deformed gauge transformations form an algebra.

\subsection{Structure of the gauge invariant EOMs in the general case}
It turns out that in the case at hand there is no need to construct interactions perturbatively because it is not difficult to describe explicitly the structure of the most general BRST differential. To begin with, we restrict our analysis to the level of equations of motion. Let $\algA$ be a linear space such that $A^I$ are coordinates on $\algA[1]$ (i.e. $\algA$ with the degree of coordinates shifted by $1$). In practice, $I$ runs over a direct sum of either $sl_2\oplus sl_2$ or $so(3,2)$ modules or just generic linear space. Consider a general gauge theory whose minimal BRST formulation involves some $A^I$ of ghost degree $1$, spacetime coordinates $x^a$ and their differentials $\theta^a$.

It follows from the general considerations~\cite{Barnich:2010sw} (see also~\cite{Grigoriev:2019ojp} for a more geometrical explanation) that BRST differential can be assumed to have the following structure:
\begin{equation}
\label{Q-full}
 Q= \theta^a\left(\dl{x^a}+ \Omega_{aI}^J(x) A^I\dl{A^J}\right)+f_{IJ}^K(x)A^I A^J \dl{A^K}\,.
\end{equation}
In fact this is a general form of a degree $1$ differential that projects to $d_X$. Explicitly, this condition reads as, $Qx^a=\theta^a$ and $Q\theta^a=0$.

The nilpotency of $Q$ implies that (1) $\Omega_{aI}^J$ are coefficients of a flat $gl(\algA)$ connection; (2) $f_{IJ}^K(x)$ are covariantly constant
\begin{align}
    \left(\dl{x^a}f_{IJ}^K(x) +\Omega_{aI}^M(x) f_{MJ}^K(x)+\Omega_{aJ}^M(x) f_{IM}^K(x) - \Omega_{aM}^K(x) f_{IJ}^M(x)\right)A^I A^J=0\,,
\end{align}
with respect to $\Omega$; (3) $f_{IJ}^K(x)$ determine a Lie algebra structure on $\algA$ for any $x$. The term linear in $\theta^a$ defines a linear gauge system. 

Suppose that $\algA$ is an $\algg$-module and $\Omega$ originates from a $\algg$-connection (this is the case for all known topological higher spin theories in 3d, $\algg$ being $sl_2\oplus sl_2$ or $so(3,2)$). Then it follows that the linearized theory determined by $Q_0= \theta^a(\dl{x^a}+ \Omega_{aI}^J A^I\dl{A^J})$ is manifestly $\algg$-invariant. Indeed, let $\epsilon^I_J(x)$ be an $\algg$-valued covariantly constant section. Then the action of $\algg$ on the BRST complex can be defined by the following $Q_0$-invariant vector field of degree $0$:
\begin{equation}
Z=\epsilon^I_J A^J\dl A_I\,.
\end{equation}
Its $Q_0$-invariance amounts to covariant constancy of $\epsilon$.
Vector field $Z$ represents the action of a global symmetry. Recall that at the level of equations of motion global symmetries are represented by ghost degree zero $Q$-invariant vector fields, while trivial symmetries correspond to $Q$-exact vector fields. The latter are the symmetries that are proportional to gauge symmetries.

Having in mind the Noether procedure, which is a perturbative approach of constructing consistent interactions starting from a free theory in a given spacetime with a symmetry algebra $\algg$, we have by default that the full interacting theory is $\algg$-invariant. If now we insist that the global $\algg$-symmetry is also a symmetry of the entire $Q$ we conclude that $f_{IJ}^K$ is an invariant tensor. Then its covariant constancy implies that it is $x$-independent. 

Let us summarize what we have learned so far: given a collection of conformal or (partially)-massless higher spin fields on AdS or Minkowski space (without matter and with the spin-one and maximal depth partially-massless fields taken to be topological), the most general interacting theory (at the level of equations of motion) that is invariant with respect to an isometry algebra $\algg$ is determined by a $\algg$-invariant Lie algebra structures on $\algA$. The associated BRST differential has the form~\eqref{Q-full}. 

A natural question is whether $Q$ arises as an expansion about some vacuum solution of a background independent theory. If this is the case the background independent theory we are talking about is the one determined by
\begin{equation}\label{diffinvq}
    Q^{\prime}= d_X+f_{IJ}^K A^I A^J \dl{A^K}\,.
\end{equation}
It is clear that for~\eqref{Q-full} to be an expansion of the above $Q^{\prime}$ about a vacuum solution it is necessary that $\algA$ contains $\algg$ as a subalgebra and that the $\algg$-module structure of $\algA$ arises from the adjoint action of $\algg$ on $\algA$.

Such strong algebraic conditions are not always satisfied so that there can be, in principle, higher spin theories that are not of Chern-Simons-type, at least at the level of equations of motion. In particular, one such example of a background-dependent theory is given in Section~\bref{subsec:hsa}. 

\subsection{Diffeomorphism invariance condition} 
Another condition that immediately forces the theory to be of Chern-Simons-type is the requirement that the full interacting theory is diffeomorphism invariant. The BRST formulation of diffeomorphism invariant theories is such that $x,\theta$-dependence factorizes. In other words, performing a local change of variables
one can bring $Q$ to the following form:
\begin{equation}
 Q=d+\bar Q\,,
\end{equation} 
where $\bar Q$ does not explicitly involve $x^a,\theta^a$.

Suppose that the interacting higher spin theory we are looking for is diffeomorphism invariant. It is then determined by some $\bar Q$ which is independent of $x^a,\theta^a$. The general form of such ghost degree 1 vector field  is
\begin{equation}
 \bar Q= A^I A^J U_{IJ}^K \dl{A^K}\,.
\end{equation} 
Let us recall that the underlying linear space $\algA$ ($A^I$ are coordinates on $\algA[1]$) is, in our case, a direct sum of a number of $sl_2\oplus sl_2 $ or $so(3,2)$ modules, associated to the fields present in the model. Tensor $U$ (now it has to be $x$-independent) determines a bilinear map $\algA\wedge \algA \to \algA$, while $Q^2=0$ ensures that this map is a Lie algebra structure on $\algA$.

To summarize, what we have arrived at is precisely the BRST differential of the Chern-Simons theory. Requiring it to be Lagrangian implies that $\algA$ is equipped with an invariant inner product which allows us to write down (BV master-) action if the spacetime dimension is 3.

\subsection{Metric like vs. frame-like vertices}
\label{sec:metric2frame}
The proper set up for cubic vertices in the metric like approach deals with BV-BRST formulation of the linear theory. The underlying space is the jet-bundle of the theory extended by ghosts and antifields. We keep denoting generating functions for fields, ghosts and their spacetime derivatives $\Phi,\Xi$, but now we only subject them to $(\d_p\cdot\d_p) \Xi=0$ and $(\d_p\cdot\d_p)^2 \Phi=0$, i.e. the transverse-traceless gauge is not assumed from the onset.

This space is equipped with the horizontal differential $\Dh=\theta^a D_a$, gauge differential $\gamma$ and Koszul-Tate differential $\delta$ so that the total differential is
\begin{equation}
\tilde s = \Dh+s_0+\delta \,.  
\end{equation}
If $\Phi^*(y,p)$, $\Xi^*(y,p)$ denote generating functions for antifields  conjugate to $\Phi$ and $\Xi$, respectively, then $\delta \Phi^*=\cF \Phi$
and $\delta \Xi^*=\Pi(\d_p\cdot \d_y) \Phi^*$, where $\cF$ defined Fronsdal action through $\inner{\Phi}{\cF\Phi}|_{y=0}$ and $\Pi$ is a projector to the kernel of $\d_p\cdot\d_p$.  Differential $s_0$ acts trivially on antifields and $\Xi$, while its action on fields is determined by $s_0\Phi=(p\cdot \d_y)\Xi$.

The cubic vertices are described by $\tilde s$ cohomology in ghost degree $d$ (spacetime dimension, in our case $d=3$). Under the usual assumptions that we are working locally in both the spacetime and in the field space the cohomology of $\Dh$ is known to be nontrivial only in degree $d$ in $\theta^a$. This implies that $\tilde s$ cohomology is isomorphic to $s=\delta+\gamma$ cohomology in the space of local functionals, i.e. local $d$-forms considered modulo $\Dh$-exact ones. More precisely, decomposing a general cocycle $V$, $\tilde s V=0$ with respect to form degree
\begin{align}
    V=V_d+V_{d-1}+\ldots +V_0
\end{align}
one finds that $V^d$ satisfies 
\begin{align}\label{descent}
(\gamma +\delta) V_d=-\Dh V_{d-1}\,,
\end{align}
which implies that antifield-independent piece of $V_d$ is gauge invariant modulo equations of motion ($\delta$-contribution) and total derivative ($\Dh$-contribution).

Other way around, given an antifield-independent $V^d_0$ that is on-shell gauge-invariant modulo a total derivative one can recursively reconstruct an $\tilde s$-cocycle $V$ using acyclicity of $\Dh$ in form degree $<d$ and acyclicity of $\delta$ in nonvanishing antifield degree. In other words, inequivalent cubic vertices are in one-to-one with $\tilde s$-cohomology in ghost degree $d$ and restricted to elements cubic in fields.

This conventional  BV-BRST complex can be equivalently reduced to a smaller complex presented in Section~\bref{BRST-Fronsdal}.  Before explaining details of the reduction let us 
recall a useful geometrical interpretation~\cite{Barnich:2004cr} of the equivalent reduction of BRST complexes of the above type. In fact this applies to generic $Q$-manifolds. The reduction can be understood as a restriction to the submanifold of the jet space which is locally determined by the equations $w^a=0$, $\tilde s w^a=0$, where variables $w^a$ are chosen in such a way that $\tilde s w^a$ are independent functions. Such variables are known as contractible pairs, while the reduction is a natural equivalence of $Q$-manifolds. It is clear that $\tilde s$ is tangent to the submanifold and hence makes an algebra of local functions on the submanifold into a homological complex.

Under certain regularity conditions this equivalence is a quasi-isomorphism of the corresponding BRST complexes (here we also disregard global geometry issues as we are focused on linear systems). The map that induces isomorphism in cohomology is simply the restriction to the submanifold. 
Upon the elimination of contractible pairs cocycle $V$ gives rise to a $Q_0$-cocycle ${V^{\prime\prime}}=V\big|_{w^a=\tilde s w^a=0}$ representing the same cohomology class. It is often referred to as a homotopy transfer of $V$. 

Let us spell out the reduction leading to the BRST complex of Section~\bref{BRST-Fronsdal} in some more details. It is convenient to split it into two steps. At the first step one eliminates contractible pairs $w^i,sw^i$, where $w^i$ are all the antifields and their spacetime derivatives (i.e. coefficients of $\Phi^*,\Xi^*$). It is easy to check that equations $w^i=0$, $\tilde s w^i=0$ also impose the equations of motion and all their derivatives, so that the reduced complex is that of functions on the stationary surface extended by ghost variables and their derivatives and the differential being $\tilde s_0=(s_0+\Dh)|_{w^i=\tilde s w^i=0}$ (sometimes it is called on-shell BRST complex).  
Upon the elimination of $w^i$, $\tilde sw^i$ cocycle $V$ gives rise to a $\tilde s_0$-cocycle ${V^{\prime}}=V\big|_{w^i=\tilde s w^i=0}$ representing the same cohomology class.

At the next step one eliminates $w^\beta$ and $\tilde s_0 w^\beta$ with $w^\beta$ being all the traces and divergences of the fields. It is easy to check that $w^\beta=0$, $\tilde s_0 w^\beta=0$ also sets to zero $(\d_y\cdot \d_y)  \Xi$ and $(\d_y\cdot \d_p) \Xi$, so that we indeed arrive at the BV-BRST formulation given in Section~\bref{BRST-Fronsdal}. 
It is then not difficult to obtain a homotopy  transfer $V^{\prime\prime}$ of $V$. According to the general prescription it is obtained by restricting $V^\prime$ to the surface determined by $w^\beta=\tilde s_0 w^\beta=0$.  

In fact it is also easy to obtain $V^{\prime\prime}$ directly from $V$. Indeed,
$V^{\prime\prime}$  coincides with $V$ where one sets to zero all the antifields as well as the components of $\Phi,\Xi$ entering $(\d_y\cdot \d_y) \Phi$, $(\d_y\cdot \d_p) \Phi$, $(\d_p\cdot \d_p) \Phi$ and  $(\d_y\cdot \d_y) \Xi$, $(\d_y\cdot \d_p) \Xi$, $(\d_p\cdot \d_p) \Xi$. In other word $V^{\prime\prime}$ is a restriction of $V$ to the surface 
\begin{equation}
\label{TT-surf}
 (\d_y\cdot \d_y) \Phi=(\d_y\cdot \d_p)\Phi =(\d_p\cdot \d_p)\Phi=0\,, \qquad  
 (\d_y\cdot \d_y) \Xi=(\d_y\cdot \d_p)\Xi =(\d_p\cdot \d_p)\Xi=0\,
\end{equation}
Note that $s_0$ restricted to this surface obviously coincides with $\gamma$ so that by construction $V^{\prime\prime}$ satisfies $Q_0 V^{\prime\prime}=0$.

The following remark is in order. If we decompose
a $Q_0$ cocycle $V^{\prime\prime}$ with respect to form degree as $V^{\prime\prime}_d+V^{\prime\prime}_{d-1}+\ldots+V^{\prime\prime}_0$ one gets
\begin{equation}
\label{on-shell-cubic}
    \gamma V^{\prime\prime}_d=-\Dh V^{\prime\prime}_{d-1}\,, \qquad \gamma V^{\prime\prime}_{d-1}=-\Dh V^{\prime\prime}_{d-2}\,, \qquad \ldots  \,, \qquad \gamma V^{\prime\prime}_{0}\,.
\end{equation}
If we lift $V^{\prime\prime}$ off the surface~\eqref{TT-surf} the first equation can be written as
\begin{equation}
    \gamma V^{\prime\prime}_d=-\Dh V^{\prime\prime}_{d-1}+A^IE_I\,,
\end{equation}
where $E_I$ denotes equations determining the surface
\eqref{TT-surf}. In other words  $V^{\prime\prime}_d$ is precisely what is called a cubic vertex in the transverse-traceless gauge.

Given such a $V^{\prime\prime}_d$ one can always reconstruct the complete $V^{\prime\prime}$ as well as an equivalent off-shell vertex
$V$. Indeed,  applying $\gamma$ to both sides of the equation \eqref{on-shell-cubic} one gets $\Dh\gamma V^{\prime\prime}_{d-1}=0$. Taking into account that $V^{\prime\prime}_{d-1}$ is a $(d-1)$-form linear in ghosts and employing the slight generalization of the known statement, see Appendix~\bref{app:TT} (see Theorem 6.3 and Corollary 6.1 of~\cite{Barnich:2000zw}) that $\Dh$ cohomology is trivial for ghost-dependent forms of form-degree $<d$ one finds that $\gamma V^{\prime\prime}_{d-1}=-\Dh V^{\prime\prime}_{d-2}$. Continuing in this way one arrives at $V^{\prime\prime}$ satisfying $(\gamma +\Dh) V^{\prime\prime}=0$ and, hence, recover the complete vertex. In Appendix~\bref{app:TT} we present a generalization of the statement from~\cite{Barnich:2000zw} to the present case and as a byproduct demonstrate  that $V^{\prime\prime}$ can be lifted to the cubic vertex in the usual Fronsdal formulation (i.e. without the transverse-traceless gauge imposed).

Let us now find a representative of $V$ in the minimal BRST complex. To arrive at this complex one eliminates further contractible pairs with $w^a$ being all the components of $\Phi$ (in the $d>3$ case one only eliminates those in the image of $p\cdot\dl{y}$). One can check that this also eliminates all the components of $\Xi$ save for those which are in the kernel of $p\cdot\dl{y}$. The resulting complex is that given in~\eqref{mincomplex}. All in all, the minimal model representative $V^{\prime\prime\prime}$ of $V^{\prime\prime}$ (and hence of the initial $V$) is obtained by setting to zero all the components of $\Phi$ as well as all the components of $\Xi$ which are in the image of $y\cdot \dl{p}$.

The crucial observation is that only the form degree $0$ term $V^{\prime\prime}_0$ may contribute to $V^{\prime\prime\prime}$. Indeed, because $V^{\prime\prime}$ is by assumption cubic in fields and ghosts and has ghost degree $3$, all $V^{\prime \prime}_i$ with $i\neq 0$ are at least linear in $\Phi$ and hence vanish after the reduction to the minimal model. Therefore, one can assume that $V^{\prime\prime \prime}= A^I A^J A^K U_{IJK}(x)$.

To see that what we are dealing with is indeed a frame-like vertex one can repeat the analysis of ~\cite{Grigoriev:2010ic,Grigoriev:2012xg}  in order to systematically reproduce the frame-like formulation using the Lagrangian parent formalism. However in the case at hand there is a concise short-cut that does not resort to Lagrangian version of the parent formalism.

Indeed, as we discussed above given a BRST complex that has the structure of a bundle over $T[1]X$ and such that the total differential projects to $d_X$ one can recover an explicit form of the equations of motion and gauge symmetries as $d_X \circ \sigma^*    =\sigma^* \circ Q_0$ where $\sigma^*$ defines a section of the bundle, i.e. in the case at hand
$\sigma^*(A^I)=A^I(x,\theta)=A^I_a(x) \theta^a$.
Similarly, gauge transformations are given by
\begin{equation}
\delta A^I(x,\theta)=d_X \epsilon^I(x)+(Q_0 A^I)|_{A^J=\epsilon^J(x)}\,.
\end{equation}
Of course, this is nothing but a frame-like formulation of the system.

The cocycle $V^{\prime\prime\prime}$ gives an on-shell gauge-invariant vertex
\begin{equation}\label{actionFLA}
\mathcal{V} (\sigma)= \int_{T[1]X} \sigma^* V^{\prime\prime\prime}\,.
\end{equation}
In terms of components this is simply
\begin{equation}\label{actionFL}
\mathcal{V}(A)=\int_{T[1]X} V^{\prime\prime\prime}|_{A^I=A^I(x,\theta)}=\int_X V^{\prime\prime\prime }(x,dx,A(x,dx))\,,
\end{equation}
where the last expression is given in the language of forms. One can check that it is indeed  gauge invariant modulo total derivatives and linearized equations of motion. Note that the above formula is a slight generalization of the natural map, known~\cite{Barnich:2009jy} (see also \cite{Vasiliev:2005zu}) in the context of AKSZ sigma models, that sends representatives of the target space cohomology classes to the field theoretical BRST cohomology of the model in the space of local functionals. This map is locally a quasi-isomorphism~\cite{Barnich:2009jy} (see also \cite{Bonavolonta:2013mza}).

Let us summarize the results of this section. Starting from the fully off-shell free higher spin theory in $3d$ (i.e. a set of free (partially)-massless or conformal higher spin fields or any other fields that have a similar structure of the minimal model), it is possible to show that
\begin{itemize}
    \item The study of interactions in a theory is equivalent to a much simpler problem of studying its minimal model; 
    
    \item It is easy to describe the BRST operator of the most general fully interacting model \eqref{Q-full}. The latter, if there is a global symmetry algebra $\mathfrak{g}$, implies that the structure functions $f_{IJ}^K$ are $x$-independent and are structure constants of a Lie algebra $\mathfrak{f}$. Provided $\mathfrak{g}$ is a subalgebra of $\mathfrak{f}$, the theory is background-independent, \eqref{diffinvq};
    
    \item The equations have the form of the flatness condition. If, in addition, there exists a $\mathfrak{f}$-invariant non-degenerate bilinear form, the equations admit an action and it is just the Chern-Simons action;
    
    \item Formula \eqref{actionFL} gives a constructive way to rewrite every metric-like interaction vertex in the frame-like language; 
    
    \item A number of important consequences of the most general formulation being of Chern-Simons-type have already been discussed in Section \bref{sec:hsgravity}. 
\end{itemize}
The main conclusion of this part is that all matter-free higher spin theories with (partially)-massless and conformal fields are of Chern-Simons type. Since the metric-like approach is equivalent to the frame-like one, this completes the Noether procedure in $3d$.

Even though in this work we are mainly interested in $3d$ higher spin theories, a considerable part of the proof is general enough as to make the following remarks about any dimension $d>2$. 
\begin{itemize}
    \item One of the general questions has always been about the equivalence between frame-like and metric-like languages. Going one way is easy: on fixing the Stueckelberg symmetries in the frame-like formulation one can solve for all frame-like fields in terms of the Fronsdal fields. Therefore, given a frame-like interaction vertex, it is always possible to rewrite it in terms of Fronsdal fields. Indeed, all the steps of the proof are applicable in general $d$ so that one can indeed find a representative $V^{\prime\prime\prime}$ of a given metric like vertex in the minimal BRST complex. But in general $d$ it also depends on degree zero variables (what remain of $\Phi$ upon elimination of the image of $p\cdot \dl{x}$). Formula \eqref{actionFLA} gives then an explicit form of the frame-like vertex. 
    
    \item The above arguments as well as the proof given in this section extends to flat space mixed-symmetry fields and to (partially)-massless (mixed-symmetry) fields in $AdS_d$ \cite{Metsaev:1997nj, Boulanger:2008up, Skvortsov:2009nv}, in which case a good starting point is the ambient space BRST complex~\cite{Barnich:2006pc, Alkalaev:2008gi,Alkalaev:2009vm,Alkalaev:2011zv}.  Note however, that in the case of mixed symmetry fields on AdS a proper Lagrangian formulation is not known in the general case so that cubic vertices can only be understood as local functions which are gauge invariant modulo total derivative on the solutions to the equations of motion.

    \item
    The advantage of the frame-like language is that it features much less structures once it comes to writing an ansatz for an interaction. Nevertheless, nothing is lost. Note that $\bar\Phi$ in \eqref{minimaleqs} does not vanish in $d>3$ and can also contribute to interaction vertices; 
    
    \item Another debatable question has been whether the transverse-traceless gauge is a restriction,\footnote{In the same spirit, one can mention \cite{Manvelyan:2010jr}, see also \cite{Fotopoulos:2010ay, Manvelyan:2010je,Mkrtchyan:2011uh,Metsaev:2012uy,Francia:2016weg} that allows one to uplift the vertices the fully off-shell covariant vertex.} i.e. whether every gauged-fixed vertex can be uplifted to a local gauge invariant vertex in the fully off-shell Fronsdal theory. We explicitly demonstrate this in Appendix \bref{app:TT}. 
\end{itemize}

\subsection{Implementation}
We would like to illustrate \eqref{actionFL} on the actual metric-like vertices in three dimensions. In practice, one begins with a cubic metric-like vertex $V^3=V^3(\Phi,\Phi,\Phi)$ in the transverse-traceless gauge. The fact that it is gauge-invariant on-shell modulo a total derivative and equations of motion (and gauge-conditions, which can be understood as a part of the equations of motion) tells that it is possible to find $V^2_\mu$ such that
\begin{align}\label{descentA}
    \delta_{\xi} V_3+ \pl_\mu V^{\mu}_2(\xi,\Phi,\Phi)&\thickapprox 0\,,
\end{align}
where $\thickapprox$ means modulo equations of motion/gauge conditions. Note that the gauge parameters here, e.g. $\xi$, are understood as anti-commuting variables (as different from the usual implementation of the Noether procedure in the non-BV-BRST language). At the second step of the descent procedure \eqref{descent} we take the variation with respect to $\xi$, which effectively yields a commutator of two gauge transformations:
\begin{align}\label{descentB}
    \delta_{\xi}  V^{\mu}_2(\xi,\Phi,\Phi)+  \pl^\nu V^{\nu\mu}_1(\xi,\xi,\Phi)&=0\,.
\end{align}
The existence of $V_1^{\nu\mu}$ follows from \eqref{descent}. At the last step we find the Jacobi identity
\begin{align}\label{descentC}
    \delta_{\xi}  V^{\nu\mu}_1(\xi,\xi,\Phi) +\pl^\rho V^{\rho\nu\mu}_0(\xi,\xi,\xi)&=0\,.
\end{align}
Now, we have $V^{\rho\nu\mu}_0(\xi,\xi,\xi)=\epsilon^{\rho\nu\mu}V_0(\xi,\xi,\xi)$. Note, that $V_{3,2,1}$ involve some derivatives that originate from those acting on $\Phi$'s in $V_3$. 

Now, we transfer $V_0$ to the minimal model to get a certain $V^{\prime\prime\prime}_0$. Let us consider massless higher spin fields in $3d$ Minkowski for definiteness. In $3d$ the transfer amounts to dropping all higher derivatives of $\xi$, i.e. $\pl^k\xi=0$, $k\geq2$. The first derivative $\pl\xi$ should be replaced in accordance with \eqref{mincomplex}, c.f. \eqref{masslesslorentz}, \eqref{framespinor}, i.e.\footnote{This should not be confused with the frame-like equations, where certain combinations of derivatives are not constrained by the equations. In the minimal model such derivatives are simply set to zero. To distinguish between $\xi^{a_1...a_{s-1}}$ and the dual of $\xi^{a_1...a_{s-1},m}$ that has the same index structure we denote the latter $\bar\xi^{a_2...a_{s-1}}$. }
\begin{align}
    \pl^m \xi^{a_1...a_{s-1}}&= \xi^{a_1...a_{s-1},m} && \Longleftrightarrow && \pl^m \xi^{a_1...a_{s-1}}=\epsilon\fud{m (a_1}{b}\bar\xi^{a_2...a_{s-1})b}
\end{align}
As a Lorentz tensor, the first derivative $\pl\xi$ contains three irreducible components: symmetric and traceless tensors of ranks $s$, $s-1$ and $s-2$. This equation implies that those corresponding to $s$, $s-2$ are set to zero, while the $(s-1)$-component should be solved for
\begin{align}
  \epsilon\fud{(a_1}{mn}\pl^m \xi^{na_2...a_{s-1})}=\bar\xi^{a_1...a_{s-1}}\,.
\end{align}
In spinorial language we have $\pl\fud{(\ga_1}{\gb}\xi^{\ga_2...\ga_{2s-2})\gb}=\bar\xi^{\ga_1...\ga_{2s-2}}$. As a result,  $V^{\prime\prime\prime}_0$ becomes a function of the coordinates $\xi$ of the minimal model. These coordinates are in one-to-one with the frame-like fields. The last step is to replace $\xi$ with $A_\mu\, dx^\mu$ to get a frame-like vertex. 

Now, let us illustrate this procedure with a number of examples. A somewhat tautological example is to begin with the Chern-Simons vertex understood as a metric-like vertex \cite{Barnich:2000zw}:
\begin{align}
    V_3&= \tfrac23 f_{IJK} A^I_\mu A^J_\nu A^K_\lambda\, \epsilon^{\mu\nu\lambda} && \delta A_\mu^I=\pl_\mu \xi^I
\end{align}
The descent equations give us in succession $V_2^\mu=-2f_{IJK} \xi^I A^J_\nu A^K_\lambda\, \epsilon^{\mu\nu\lambda}$, $V_1^{\nu\mu}=2f_{IJK} \xi^I \xi^J A^K_\lambda\, \epsilon^{\mu\nu\lambda}$ and $V_0^{\lambda\nu\mu}=-\tfrac23f_{IJK} \xi^I \xi^J \xi^K\, \epsilon^{\mu\nu\lambda}$. This clearly gives back $\tfrac23f_{IJK} A^I\wedge A^J\wedge A^K$.

More interesting example is the two-derivative spin-two vertex. It is convenient to write it in the language of generating functions, see in Appendix \bref{app:genfunc} for detail. We begin with $V_3$ that has two derivatives and three spin-two fields $\Phi^{ab}$. At each step of the descent one derivative is added via the gauge transformations $\delta \Phi_{ab}=\pl_a\xi_b+\pl_b\xi_a$ and one derivative is removed in each of \eqref{descentA}-\eqref{descentC}. Therefore, we get $V_0$ that is trilinear in $\xi^a$ and has two derivatives. Going to the minimal model, we set $\pl_a\pl_b\xi^c=0$, $\pl_a\xi_b+\pl_b\xi_a=0$ and $\pl_a\xi_b-\pl_b\xi_a=\xi_{ab}$. As a result, we end up with a unique expression
\begin{align}
    V^{\prime\prime\prime}_0&= \xi^a \,\xi\fud{b}{m}\,\xi^{cm}\epsilon_{abc}
\end{align}
that is then mapped to the frame-like Einstein-Hilbert vertex $e^a\wedge  \omega\fud{b,}{m}\wedge \omega^{c,m}\epsilon_{abc}$. 

It is easy to consider the most general case of odd/even interaction vertex. The crucial advantage of $3d$ is that there is a unique such vertex and, moreover, there is a unique expression\footnote{In $d>3$ there is also a unique interaction vertex in constant-curvature space-times with given spins and a number of derivatives (or none at all). However, there is no unique expression for $V^{\prime\prime\prime}_0$. Therefore, a computation needs to be done to get the frame-like vertex explicitly. } that can serve as $V^{'''}_0$. This is due to the fact that the Chern-Simons vertex is unique, while usually vertices are classes of equivalence modulo field redefinitions. As a result, in Minkowski space the vertex with two derivatives gives 
\begin{align}
    e^{a u(k)v(n)} \wedge \omega\fud{b}{u(k)w(m)} \wedge \omega\fud{cw(m)}{v(n)}\epsilon_{abc}
\end{align}
and the vertex with three derivatives 
\begin{align}
    \omega^{a u(k)v(n)} \wedge \omega\fud{b}{u(k)w(m)} \wedge \omega\fud{cw(m)}{v(n)}\epsilon_{abc}\,,
\end{align}
where $s_1=k+n+2$, $s_2=m+k+2$, $s_3=m+n+2$. The details on how to do it are given in Appendix \bref{app:genfunc}. We checked it on several examples.

\section*{Acknowledgments}
\label{sec:Aknowledgements}
We are grateful to Glenn Barnich and Stefan Fredenhagen for useful discussions. The work of M.G. and E.S. was supported in part by the Russian Science Foundation grant 18-72-10123 in association with the Lebedev Physical Institute. The work of KM was supported in part by Scuola Normale, by INFN (IS GSS-Pi) and by the MIUR-PRIN contract 2017CC72MK\_003.

\begin{appendix}
\renewcommand{\thesection}{\Alph{section}}
\renewcommand{\theequation}{\Alph{section}.\arabic{equation}}
\setcounter{equation}{0}\setcounter{section}{0}

\section{From TT-gauge to off-shell vertices}
\label{app:TT}

In this appendix, we suggest a method to uplift the TT vertices to full off-shell, concentrating on the example of Fronsdal fields in Minkowski space, for which the off-shell completions are already known \cite{Manvelyan:2010jr}, see also \cite{Fotopoulos:2010ay, Manvelyan:2010je,Mkrtchyan:2011uh,Francia:2016weg}.

In the setting of Section~\bref{sec:metric2frame}
let us restrict ourselves to Fronsdal fields in Minkowski space but keep the spacetime dimension $d$ generic ($d>2$). Suppose that $V_d^{\prime\prime}$ is a vertex in the transverse-traceless gauge, i.e. is a $d$-form cubic in $\Phi$ defined on the surface $\d_y\cdot\d_y \Phi=\d_y\cdot\d_p \Phi=\d_p\cdot\d_p \Phi=0$ (but we lift it to functions defined off the surface) and satisfying   $\gamma V_d^{\prime\prime}+\Dh V_{d-1}^{\prime\prime}=0$ modulo terms vanishing on the surface.

A technical trick to show that $V^{\prime\prime}_{d}$ can be lifted to an off-shell vertex in the Fronsdal formulation and to a complete on-shell cocycle of $\Dh+\gamma$ is to employ a resolution of the surface using a suitable differential which can be thought of as an extension of the usual Koszul-Tate differential. To this end we introduce generating function $\Psi(y,p,b,c_0,c,c_T)$ for fields and antifields, where $c_0,c_+,c_T$ are fermionic ghost variables of degree $1$ and $b$ of degree $-1$. The ghost degree and Grassmann parity of the components are set by requiring $\gh{\Psi}=\p{\Psi}=0$. In particular, generating functions $\Phi(y,p)$ and $\Xi(y,p)$ are identified with the ghost-independent component and the linear in $b$ component of $\Psi$ respectively. Note that in addition to $\Phi$ there are further components of degree $0$ which enter $\Psi$ as terms linear in $b$ and $c$-ghosts.

Now Koszul-Tate-like differential is defined by
\begin{equation}
\delta^{K}\Psi=\Omega_{-1} \Psi\,, \qquad \Omega_{-1}=c_0(\d_y\cdot \d_y)+c(\d_p\cdot\d_y)+c_T(\d_p\cdot\d_p)\,,
\end{equation}
where $\delta^{K}$ is a vector field acting on components, while $\Omega$ acts on auxiliary variables $y,p,c_0,c,c_T$. Operator $\Omega$ was employed in studying BRST complex for Fronsdal fields in~\cite{Barnich:2004cr}, where it was shown to have cohomology only in vanishing degree in $c_0,c,c_T$. In terms of $\delta^{K}$ this implies that cohomology of $\delta^{K}$ is concentrated in the vanishing resolution degree and are precisely given by functions of $\Phi(y,p)$ and $\Xi(y,p)$ restricted to the surface~\eqref{TT-surf}. The resolution degree is induced by the homogeneity in $c_0,c,c_T$, e.g. $\mathrm{rdeg}(\Phi)=\mathrm{rdeg}(\Xi)=0$ and $\mathrm{rdeg}(\delta^K)=-1$.  To see that cohomology indeed coincides with the functions of $\Phi(y,p)$ and $\Xi(y,p)$ restricted to the surface~\eqref{TT-surf} 
one observes that $\delta^{K}$-exact functions in degree $0$ are necessarily proportional to the RHS $\d_y\cdot\d_y \Phi$, $\d_y\cdot\d_p \Phi$, $\d_p\cdot\d_p \Phi$ of the equations, \cite{Barnich:2004cr}, or to analogous constraints with $\Xi$.

Let us now show that given $V_d^{\prime\prime}$ satisfying $\gamma V_d^{\prime\prime} +\Dh V_{d-1}^{\prime\prime}+\delta^{K} V^{\prime\prime}_{d|1}=0$ with some $V_{d|1}$ of resolution degree $1$ (this is just a reformulation of the on-shell gauge invariance condition in terms of $\delta^{K}$) one can construct $W$, $\gh{W}=d$ depending also on antifields and ghosts such that $(\delta^{K}+\gamma+\Dh)W=0$ and such that its form degree $d$ and antifield degree $0$ component $W_{d,0}$ coincides with $V^{\prime\prime}_d$. Such $W$ can be constructed recursively using as an auxiliary degree $(d-\text{form degree}+\text{resolution degree})$ satisfying $\mathrm{adeg}(\gamma=0),\mathrm{adeg}(\Dh+\gamma)=-1$ so that $\mathrm{adeg}(V_d^{\prime\prime})=0$ and $\mathrm{adeg}(V_{d-1}^{\prime\prime})=1$
$\mathrm{adeg}(V_{d|1}^{\prime\prime})=1$. Indeed, if cohomology of $\Dh+\gamma$ is trivial in auxiliary degree $>0$ the full $W$ can be reconstructed. This can be equivalently phrased as triviality of the cohomology $H(\delta^E|\Dh)$
($\delta^E$ modulo $\Dh$) in auxiliary degree $>0$. This statement was proved in~\cite{Barnich:2000zw} for elements at least linear in ghosts and in the case where $\delta^{K}$ is a conventional Koszul-Tate differential. However, only the triviality of $\delta^K$ cohomology in nonzero resolution degree  is crucial in the proof so it extends to the $\delta^K$ above.

Let us now discuss an interpretation of the $\Dh+\gamma+\delta^K$-cocycle $W$ in the extended system. Introducing differential $s=\gamma+\delta^K$ it is easy to see that in terms of generating function $\Psi$ it is determined by 
\begin{equation}
s\Psi=\Omega  \Psi\,, \qquad \Omega=(p\cdot\d_y)\dl{b}+c_0(\d_y\cdot \d_y)+c(\d_p\cdot\d_y)+c_T(\d_p\cdot\d_p)
+\ldots\,,
\end{equation}
where $\ldots$ denote the ghost term encoding the constraint algebra. This is precisely the BRST operator  of the so-called triplet system, where the trace constraint is incorporated in the BRST operator (see \cite{Barnich:2004cr} for detailed discussion). 

In this way we proved that any cubic vertex in the transverse-traceless gauge can be lifted to that in the extended triplet formulation. Although this formulation is not manifestly Lagrangian it can be equivalently reduced~\cite{Barnich:2004cr} (by eliminating contractible pairs for the term in $\Omega$ proportional to $c_T$) to the Lagrangian formulation. The reduced formulation BV master action is given  
\begin{equation}
S_{BV}=\int d^dx\, \inner{\Psi}{\Omega_0\Psi}\,, \qquad \Omega_0=(p\cdot\d_y)\dl{b}+c_0(\d_y\cdot\d_y)+c(\d_p\cdot\d_y)-c\dl{b}\dl{c^0}\,,
\end{equation}
and where $\Psi$ is subject to $\dl{c^T}\Psi=0=(\d_p\cdot \d_p-2\dl{b}\dl{c})\Psi$.  This is a usual triplet form of the free higher spin theory. In its turn this formulation gives conventional Fronsdal one through elimination of the auxiliary field $C$ entering $\Psi$ as a coefficient of $c_0b$. The $d$-form component of the vertex $W$ (with $C$ eliminated) is clearly a usual off-shell cubic vertex of the Fronsdal system.

\section{Cubic Vertices}
\label{app:genfunc}
\setcounter{equation}{0}
In three dimensions, vertices for arbitrary spins have a relatively simple form. One subtle point is that one needs to take care of Schouten identities. In order to set up notations, we introduce the fields $\phi^{(s_i)}(x_i,a_i)=\phi^{(s_i)}_{\mu_1\dots\mu_{s_i}}(x_i)a_i^{\mu_1}\dots a_i^{\mu_{s_i}}$, $i=1,2,3$, and write the vertex in the form:
\begin{equation}
    V_{s_1,s_2,s_3}=\mathcal{V}(P,A)\phi(x_1,a_1)\phi(x_2,a_2)\phi(x_3,a_3)\vert_{a_i=0, x_i=x}\,,
\end{equation}
where the vertex operator depends on $P_i^{\mu}=\partial^{\mu}_{x_i}, A_i^{\mu}=\partial^{\mu}_{a_i}$. We also introduce the notion of total derivative: $P^\mu=P_1^{\mu}+P_2^{\mu}+P_3^{\mu}$. The vertex operator depends on the twenty-one elementary scalar contractions $P_i\cdot P_j$, $P_i\cdot A_j$ and $A_i\cdot A_j$, among which nine are trivial on-shell:
\begin{align}
    P_i^2=0\,,\quad P_i\cdot A_i=0\,,\quad A_i^2=0\,, \quad (i=1,2,3)\,,
\end{align}
and six:
\begin{equation}
    P\cdot P_i\,,\quad P\cdot A_i\,,\quad (i=1,2,3)
\end{equation}
form total derivatives. The remaining six variables can be given by:
\begin{align}
    y_i=A_i\cdot P_{i+1}\,, \quad z_i=A_{i+1}\cdot A_{i-1}\,,
\end{align}
which form elementary building blocks of the traceless-transverse (TT) part of parity-preserving cubic vertices: 
\begin{equation}
    V_{s_1,s_2,s_3}=\mathcal{V}(y_i,z_i)\phi(x_1,a_1)\phi(x_2,a_2)\phi(x_3,a_3)\Big|_{a_i=0\,, x_i=x}\,.
\end{equation}
For simplicity, we will start from parity-even vertices. In the frame-like language they will require parity odd-structures.
We therefore also introduce parity-odd structures in three dimensions \cite{Kessel:2018ugi}:
\begin{align}
    u=\epsilon_{\mu\nu\rho}A_1^{\mu}A_2^{\nu}A_3^{\rho}\,,\quad v_{ij}=\epsilon_{\mu\nu\rho}A_{i+1}^{\mu}A_{i-1}^{\nu}P_{j}^{\rho}\,, \quad w_i=\epsilon_{\mu\nu\rho}A_i^{\mu}P_{i+1}^{\nu}P_{i-1}^{\rho}\,,
\end{align}
as well as parity-odd total-derivative structures:
\begin{align}
    \tilde{v}_i=\epsilon_{\mu\nu\rho}A_{i+1}^\mu A_{i-1}^\nu P^\rho=\sum_{j=1}^3 v_{ij}\,, \quad \tilde{w}_{ij}=\epsilon_{\mu\nu\rho}A_i^{\mu}P_j^\nu\,P^\rho\,,\quad x=\epsilon_{\mu\nu\rho}P^\mu P_i^\nu P_{i+1}^\rho\;\, (\forall i)\,,
\end{align}
where $\sum_{j=1}^3 w_{ij}=0\,$. Since the $v_{ij}, \tilde{v}_i$ are redundant, we choose $v_{ij}$ with $j\neq i$ as independent variables and express the vertices in terms of them as in \cite{Kessel:2018ugi}. One can also make use of generating functions of higher-spin fields
\begin{align}
    \Phi(a,x)=\sum_{s=0}^{\infty} \frac1{s!}\phi^{(s)}(a,x)\,,
\end{align}
to write a generating functions for cubic vertices involving all triplets $s_1,s_2,s_3$.
We first write the general form of the gauge transformations for any spin:
\begin{align}
    \delta^{(0)} \phi^{(s)}(a,x)=a\cdot \partial_x\, \epsilon^{(s-1)}(a,x)\,,
\end{align}
or, in the generating function form,
\begin{align}
    \delta^{(0)} \Phi(a,x)&=a\cdot \partial_x\, \Lambda(a,x)\,, &&
    \Lambda(a,x)=\sum_{k=0}^{\infty} \frac1{k!}\epsilon^{(k)}(a,x)\,.
\end{align}
It is worth noting  that, while the gauge parameter has one lower rank compared to the gauge field for each given spin, in the generating function form involving infinite number of fields, it is ``in the same class of functions as the field itself''.

The cubic vertices are given in \cite{Mkrtchyan:2017ixk,Kessel:2018ugi} and are gauge invariant up to total derivatives. In deriving these vertices we discard boundary terms. Here, we will need exactly these boundary terms.
Note the useful (on-shell) relation:
\begin{align}
    P_{i+1}\cdot P_{i-1}=\frac12(P\cdot P_{i+1}+P\cdot P_{i-1}-P\cdot P_i)=\frac12 P^2-P\cdot P_i\,.
\end{align}
We will discard all of the on-shell trivial terms in the cubic action and its variations, but keep total derivatives, therefore, will work with functions of variables $y_i, z_i, P\cdot P_i, P\cdot A_i, v_i, w_{ij}$.
We note here, that we will be interested in at most three total derivatives. While taking gauge variations of the vertex functions, we will strip off the total derivatives and contract the free index with a Grassmann vector variable $\zeta^{\mu}$ in the following sense ($\delta_i^{(0)}$ is the lowest-order gauge variation of the $i$-th field):
\begin{align}
\delta_1^{(0)}\delta_2^{(0)}\delta_3^{(0)}\mathcal{V}^3(y,z)=\mathcal{V}^0(\zeta\cdot P_i,\zeta\cdot A_i,y_i,z_i,u,v_{ij},w_i,\bar{v}_i,\bar{w}_{ij},\bar{x}_i)\,,\\
\mathcal{V}^{(0)}(y,z,u,v)=\frac16 \epsilon_{\mu\nu\rho}\frac{\partial}{\partial \zeta_\mu}\frac{\partial}{\partial \zeta_\nu}\frac{\partial}{\partial \zeta_\rho}\mathcal{V}^{0}(\zeta\cdot P_i,\zeta \cdot A_i,y_i,z_i,u,v_{ij},w_i,\bar{v}_i,\bar{w}_{ij},\bar{x}_i)\,,\label{FrameVertexSeed}
\end{align}
where
\begin{align*}
\frac{\partial}{\partial \zeta_{\mu}}=A_i^\mu \, \frac{\partial}{\partial (\zeta\cdot A_i)}+P_i^\mu \, \frac{\partial}{\partial (\zeta\cdot P_i)}+\epsilon_{\mu\nu\rho}A_{i+1}^{\nu}A_{i-1}^{\rho}\frac{\partial}{\partial \bar{v}_i}+\epsilon_{\mu\nu\rho}A_i^{\nu} P_j^{\rho}\frac{\partial}{\partial \bar{w}_{ij}}+\epsilon_{\mu\nu\rho}P_{i+1}^\nu P_{i-1}^\rho\frac{\partial}{\partial \bar{x}_i}\,,\label{dth}\\
\bar{v}_i=\epsilon_{\mu\nu\rho}A_{i+1}^\mu A_{i-1}^{\nu}\zeta^\rho\,,\quad \bar{w}_{ij}=\epsilon_{\mu\nu\rho}A_i^\mu\,P_j^{\nu}\,\zeta^{\rho}\,,\quad \bar{x}_i=\epsilon_{\mu\nu\rho}\zeta^{\mu}P^{\nu}_{i+1} P_{i-1}^{\rho}\,,
\end{align*}
and we do sum over $i,j$ indices in all the terms above. 
The expression $\mathcal{V}^{(0)}$ defines the frame-like vertex and we will see later that it will not depend on the parity-odd structures $w_i$. 

Note, that we work with the vertex operators symbolically, assuming the antisymmetry of the underlying gauge parameters (ghosts) they act on. Otherwise, the expression \eqref{FrameVertexSeed} would be trivial.

In order to proceed to gauge transformations of the vertex, we remind that in the process of passing from a metric-like vertex to the frame-like one, we will need to keep the total derivatives, therefore, we are dealing with vertex operators $\mathcal{V}^{n}$ that contain the Grassmann vectors $\zeta$ replacing the total derivative operator $P_{\mu}$ in $(3-n)$ structures $\zeta\cdot P_i, \zeta\cdot A_i$. Also, at the very end of the procedure we strip off all the three operators $\zeta_\mu$ and multiply the resulting third rank tensor with the fully antisymmetric tensor $\epsilon_{\mu\nu\rho}$, therefore any expression with symmetrized indices can be assumed to vanish: $\zeta_{(\mu}\zeta_{\nu)}=0$, thus we can take the $\zeta$'s to be Grassmannian to automatically satisfy this condition. In particular, 
\begin{equation}
\zeta^2=0\,,\quad (\zeta\cdot P_i)^2=0=(\zeta \cdot A_i)^2\,,\quad \zeta\cdot P=0\,.
\label{th}
\end{equation}
The last equation drops total derivatives in the final frame-like vertex seed \eqref{FrameVertexSeed}.
It is also straightforward to show using \eqref{th}, that,
\begin{align}
    \zeta\cdot P_i\; \zeta\cdot P_j=0\,.
\end{align}
The latter equation implies that the frame-like vertex does not depend on $w_i$, as mentioned above. Parity-odd structures $v_{ij}$ with $j\neq i$ are curl operators contracted with an index from another field via $\epsilon$-tensor, thus have a simple interpretation in the frame-like language. The operator $u$ is a contraction of one index from each of the three fields with an $\epsilon$-tensor and is naturally translated to frame language.

The gauge transformation acts on the vertex operator in the following form:
\begin{align}
    D_i =\zeta\cdot P_i\, \frac{\partial}{\partial\, (\zeta\cdot A_i)}+(\zeta\cdot A_{i+1}-y_{i+1})\frac{\partial}{\partial z_{i-1}} -\zeta\cdot P_{i-1}\,\frac{\partial}{\partial y_i} +y_{i-1} \frac{\partial}{\partial z_{i+1}}+\bar{x}_i\frac{\partial}{\partial w_i}\nonumber\\
    +(\tilde{w}_{i-1 i}-w_{i-1})\frac{\partial}{\partial v_{i+1 i-1}}-(\tilde{w}_{i+1 i}+w_{i+1})\frac{\partial}{\partial v_{i-1 i+1}}+(\bar{v}_{i}-v_{i i+1}-v_{i i-1})\frac{\partial}{\partial u}\,.\label{Di}
\end{align}

\end{appendix}

\footnotesize
\providecommand{\href}[2]{#2}\begingroup\raggedright\endgroup

\end{document}

%% file: YoungDiagrams.tex
\newcommand{\bep}{\begin{picture}}
\newcommand{\eep}{\end{picture}}

\newcounter{YoungHeight}\newcounter{YoungWidth}

\newcounter{Mul1}\newcounter{Mul2}\newcounter{Mul3}\newcounter{Mul4}
\newcounter{A0}\newcounter{A1}\newcounter{A2}
\newcounter{B3}
\newcounter{C3}\newcounter{C4}
\newcounter{D1}\newcounter{D2}\newcounter{D3}
\newcounter{T0}\newcounter{T1}

\newlength{\txtHShift}

\newlength{\txtWidth}

\newcommand{\HalfLength}[2]{\setcounter{Mul1}{#1}\setcounter{Mul2}{#1}\addtocounter{Mul1}{\value{Mul2}}\addtocounter{Mul1}{\value{Mul2}}%
\addtocounter{Mul1}{\value{Mul2}}\addtocounter{Mul1}{\value{Mul2}}\setcounter{#2}{\value{Mul1}}}

\newcommand{\Add}[3]{\setcounter{#1}{#2}\addtocounter{#1}{#3}}

\newcommand{\Length}[1]{#10}
\newcommand{\YoungScale}{}

\newcommand{\shiftedText}[2]{{\hspace{#1}#2}}
\newcommand{\calcHShift}[1]{\settowidth{\txtWidth}{#1}\setlength{\txtHShift}{-0.5\txtWidth}}

\newcommand{\TextTop}[3]{{\calcHShift{#1}\HalfLength{#2}{T0}\Add{T1}{\Length{#3}}{-9}\put(\value{T0},\value{T1}){\shiftedText{\txtHShift}{#1}}}}

\newcommand{\BlockA}[2]{{\YoungScale\bep(\Length{#1},\Length{#2}){\Add{A1}{#1}{1}\Add{A2}{#2}{1}}%
\multiput(0,0)(10,0){\value{A1}}{\line(0,1){\Length{#2}}}\multiput(0,0)(0,10){\value{A2}}{\line(1,0){\Length{#1}}}%
\setcounter{YoungHeight}{\Length{#2}}\setcounter{YoungWidth}{\Length{#1}}\eep}}

\newcommand{\RectT}[3]{\bep(\Length{#1},\Length{#2})\put(0,0){\line(1,0){\Length{#1}}}\put(0,0){\line(0,1){\Length{#2}}}%
\put(\Length{#1},\Length{#2}){\line(-1,0){\Length{#1}}}\put(\Length{#1},\Length{#2}){\line(0,-1){\Length{#2}}}#3{#1}{#2}\eep}

\newcommand{\RectBRow}[4]{{\bep(\Length{#1},20)\put(0,0){\RectT{#2}{1}{\TextTop{#4}}}%
\put(0,10){\RectT{#1}{1}{\TextTop{#3}}}\eep}}

\newcommand{\YoungA}{\BlockA{1}{1}}

\newcommand{\YoungAA}{\BlockA{1}{2}}

\newcommand{\BlockApar}[2]{\parbox{\Length{#1}pt}{\YoungScale\bep(\Length{#1},\Length{#2}){\Add{A1}{#1}{1}\Add{A2}{#2}{1}}%
\multiput(0,0)(10,0){\value{A1}}{\line(0,1){\Length{#2}}}\multiput(0,0)(0,10){\value{A2}}{\line(1,0){\Length{#1}}}%
\setcounter{YoungHeight}{\Length{#2}}\setcounter{YoungWidth}{\Length{#1}}\eep}}

\newcommand{\BlockBpar}[4]{\parbox{\Length{#1}pt}{\YoungScale\Add{B3}{\Length{#2}}{\Length{#4}}%
\bep(\Length{#1},\value{B3})\put(0,\Length{#4}){\BlockA{#1}{#2}}%
\put(0,0){\BlockA{#3}{#4}}\setcounter{YoungHeight}{\value{B3}}\setcounter{YoungWidth}{\Length{#1}}\eep}}

\newcommand{\YoungpA}{\BlockApar{1}{1}}
\newcommand{\YoungpB}{\BlockApar{2}{1}}

\newcommand{\YoungpD}{\BlockApar{4}{1}}

\newcommand{\YoungpAA}{\BlockApar{1}{2}}

\newcommand{\YoungpBB}{\BlockApar{2}{2}}
\newcommand{\YoungpCA}{\BlockBpar{3}{1}{1}{1}}

